\documentclass{ws-ijmpa}
\usepackage{graphicx}
\usepackage{hyperref}
\usepackage{xcolor}

\usepackage[sort&compress,numbers
]{natbib}

\newcommand{\diff}{\mathrm{d}}
\renewcommand{\l}{\left(}
\renewcommand{\r}{\right)}
\renewcommand{\d}{\partial}
\newcommand{\eV}{\mathrm{\,eV}}
\newcommand{\keV}{\mathrm{\,keV}}

\begin{document}

\title{Refining lower bounds on sterile neutrino dark matter mass from estimates of phase space densities in dwarf galaxies}

\author{Fedor Bezrukov}
\address{Department of Microbiology and Molecular Medicine, University of Geneva,\\
    1211 Geneva, Switzerland}

\author{Dmitry Gorbunov}
\address{Institute for Nuclear Research of the Russian Academy of Sciences,\\
    60th October Anniversary prospect, 7a, Moscow 117312, Russia}
\address{Moscow Institute of Physics and Technology,\\
    Institutsky lane 9, Dolgoprudny, Moscow region, 141700, Russia}

\author{Ekaterina Koreshkova}
\address{Department of Particle Physics and Cosmology, Physics Faculty, M.V. Lomonosov Moscow State University,\\
    Vorobjevy Gory, 119991 Moscow, Russia}
\address{Institute for Nuclear Research of the Russian Academy of Sciences,\\
    60th October Anniversary prospect, 7a, Moscow 117312, Russia}

\maketitle

\begin{history}
\received{Day Month Year}
\revised{Day Month Year}
\end{history}

\begin{abstract}
    Dwarf spheroidal galaxies (dSphs) are recognized as being highly dominated by Dark Matter (DM), making them excellent targets for testing DM models through astrophysical observations. One effective method involves estimating the coarse-grained phase-space density (PSD) of the galactic DM component. By comparing this PSD with that of DM particles produced in the early Universe, it is possible to establish lower bounds on the DM particle mass. These constraints are particularly relevant for models of warm DM, such as those involving sterile neutrinos. Utilizing the GravSphere code, we obtain a fit of the DM PSD based on the latest reliable stellar dynamics data for twenty of the darkest dSphs, refining earlier lower bounds on sterile neutrino masses in non-resonant production scenarios.
    Additionally, we introduce an alternative approach involving the Excess Mass Function (EMF), which yields even tighter constraints. Specifically, using the maximum PSD, we derive a lower bound of 
    $m>1.02$\,keV  
    at 95\% confidence level, while the EMF method provides a stronger limit of  
    $m>1.98$\,keV
    at 95\%~CL. For the general thermal relic fermion dark matter mass the limits translate into $m>0.28$\,keV and $m>0.49$\,keV, respectively.
    Both methods are versatile and can be extended to more complex DM production mechanisms in the early Universe.
    For the first time, we also constrain parameters of models involving non-standard cosmologies during the epoch of neutrino production. Our analysis yields $m>2.54$\,keV for models with kination domination and $m>4.71$\,keV for scenarios with extremely low reheating temperature.

\end{abstract}

\ccode{PACS numbers:}


\section{Introduction}

DM phenomena are widely accepted to be the most convincing 
among indirect evidences for incompleteness of the Standard Model of
particle physics (SM). Any stable (at cosmological time scale),
electrically neutral, and almost collisionless particle, produced in the
early Universe and decoupled from primordial plasma well before
recombination, may serve as DM. Many physically
well-motivated extensions of the SM provide candidates for DM, with variety of masses and couplings. These
parameters (related directly or indirectly to DM production in the early
Universe) determine possible strategies to probe the models by means 
alternative to astronomical observations. Extensive searches for various DM
particle hints are undertaken in deep underground, under-water and
under-ice laboratories, at colliders, by ground telescopes, and by
satellites.

So far, however, only astronomical observations 
analysed  within General Relativity and SM 
provide the most direct evidence for the DM
hypothesis at late cosmological epochs pointing at the lack of
gravitational potentials at various spatial scales. Remarkably, detailed information about the missing
gravitational potentials can be used in testing the DM models. Namely, the
extraction of the missing mass density profile and measurement 
of star velocity dispersion
give an estimate 
of the DM PSD which,
  for collisionless component, remains constant according to the
  Liouville theorem. Actually, exact determination of DM density
  and velocity profiles would allow one to
  distinguish between the DM models and finally pin down those which
  fit the data. However, in practice, real observations  
always imply some averaging, so that the final quantity extracted
from the observations is the \emph{coarse grained} PSD. This quantity decreases with time, therefore its maximum provides a
\emph{lower bound} on the maximum PSD of DM
particles. The latter depends on the DM model, hence in this way 
the DM models can be explored. 

The obtained constraint is most relevant for models with noticeable velocity of DM
particles \cite{deVega:2009ku}, as they predict smaller PSD. This class of
models is called \emph{Warm} DM (WDM), and is loosely defined as models
where DM particle velocity is about $10^{-3}$ at the time of 
matter-radiation equality, when 
the relic plasma temperature was about
$T_{eq}\approx0.8$\,eV. As a physically motivated example (for others see e.g.\,\cite{Rajagopal:1990yx,Gorbunov:2008ui,King:2012wg}) we consider models with sterile neutrinos: they are produced in primordial plasma via oscillations of active neutrinos, and in keV-mass range naturally become WDM, see e.g.\,\cite{Kusenko:2009up,Merle:2013gea,Drewes:2016upu,Abazajian:2017tcc,Boyarsky:2018tvu}.

Naturally, the best data to confront the
DM model predictions are provided by the dSphs, the most compact, cold, and
DM dominated cosmological structures we are aware about. The observations of dSphs
have been used to constrain the DM models, in particular, sterile neutrinos, see e.g.\,\cite{Boyarsky:2008ju,Gorbunov:2008ka,Horiuchi:2013noa,Wang:2017hof,Alvey:2020xsk}.    
In this paper we follow and further develop the approaches of Refs.\,\cite{Gorbunov:2008ui,Alvey:2020xsk} and use the most recent and reliable observations of dSphs 
to refine the previously obtained constraints 
on non-resonant sterile neutrino productions via oscillations in the lepton-symmetric plasma of the early Universe. 
Then we consider the EMF \cite{Kaplinghat:2005sy}
as an alternative to the estimate based on the maximum of the coarse grained PSD. We observe, that it generally provides stronger lower bounds on the DM particle mass. 

Determination of DM PSD requires knowledge of DM particle velocities, non-detectable by astronomical observations. Hence, we must involve some assumption here. A naive one is that the DM velocity dispersion coincides with that of stars given both dynamics are governed by the same gravitational force. However, one may argue that we certainly know many structures, where this assumption does not work: e.g.\ all the disk galaxies exhibit quite a different pattern. Instead, one may suggest that velocity dispersion of DM particles is 3-dimensionally symmetric (isotropic), in which case it can be entirely inferred from the DM density profile. However, observations of stars indicate this is certainly not the case for the visible matter. Alternatively one may suggest an anisotropic velocity dispersion with parameters distributed in accordance with what numerical simulations of artificial DM particles forming a halo exhibit. In our study we try to use all of the three ideas to some extent and discuss the differences and possible improvements in the future.    

We start our analysis by taking the estimates of average one-dimension star velocity dispersions and galaxy core sizes (as well as the suggested errors) for a set of dSphs from literature\,\cite{Munoz:2018}, assume the DM velocities are similar to stellar ones, and calculating PSD directly from them. Estimating the maximum of PSD and EMF obtained in this way we compare them with predictions from the non-resonant production of sterile neutrino DM. We estimate both the central values and error bars of the sterile neutrino mass and find that they vary too much from galaxy to galaxy: even  taking the error bars into account, they differ by up to two orders of magnitude. We find this approach unreliable to place a lower limit on the DM particle mass. 

In our further analysis,
following the approach of Ref.\,\cite{Alvey:2020xsk}, we perform a fit to the currently available observations of galaxy star dynamics and infer PSDs to constrain the sterile neutrino mass in the models with non-resonant sterile neutrino DM production\,\cite{Dodelson:1993je}. For the first time we present the limits from both coarse-grained maximum PSD estimate and EMF. Finally, we apply the procedure to alternative DM spectra, predicted for the same oscillation production mechanism but operating in non-standard cosmological models: one with kination domination at production and another with very low reheating temperature in the early Universe, see Ref.\,\cite{Gelmini:2019wfp} for details. In both cases the obtained limits for each galaxy are found to be stronger than that in the case of standard cosmology. 

The paper is organized as follows. Section\,\ref{sec:General} 
briefly describes the quantity we are dealing with---the coarse grained
PSD---and its relation with DM momentum distribution 
as we have it after DM decoupling from plasma. In practical observations we deal with the coarse-grained quantities, and extract the inequalities for two observables related to PSD: maximum of PSD and EMF.  
In Section \ref{sec:analytical} we adopt the direct astronomical estimates of the stellar density and velocities, get analytic expression for the DM density and assume DM velocities to be similar to those of stars to arrive at the analytic lower limits for the DM particle mass. In Section \ref{sec:numerical} we apply  GravSphere 
code\,\footnote{\texttt{https://github.com/justinread/gravsphere}}  to observational stellar data and  reconstruct detailed DM density profile and reconstruct its velocity dispersion constraining their asphericity from existing DM simulations. It allows us to obtain more reliable bounds. Alternative spectra of sterile neutrinos produced in the early Universe are considered in Section\,\ref{sec:alternatives}. We conclude Section\,\ref{sec:Concl} by comparison of our results with those in Ref.\,\cite{Alvey:2020xsk} and 
discussion of prospects in hunting for DM with future observations of dSphs.

\section{Phase space density and realistic observables}
\label{sec:General}

The Liouville theorem states that the PSD of the component which does not participate in contact interactions remains intact. It implies that the DM PSD $F({\bf x},{\bf p},t)$ after decoupling (from the primordial plasma or any other source) in the early Universe is conserved in time, i.e.\ obeys the following equation
\[
\frac{\d F}{\d t}+H\l {\bf x}\frac{\d F}{\d {\bf x}} - {\bf p} \frac{\d F}{\d {\bf p}}\r=0\,,
\]
where $H=d\ln(a)/dt$ is the Hubble parameter, and $a$ in the cosmic scale factor. 

Later the DM particles form the cosmic large scale structure. Galaxy masses are then dominated by the DM, with baryonic component being almost negligible in the dark dSph. Inside a galaxy with Newtonian gravitational 
potential $\phi$ formed by the DM particles the same Liouville equation may be written as 
\[
\frac{\d F}{\d t}+\frac{\bf p}{m}\frac{\d F}{\d {\bf x}} - mG\frac{\d \phi}{\d {\bf x}} \frac{\d F}{\d {\bf p}}=0,
\]
where $G$ is the Newtonian gravitational constant and $m$ is the mass of the DM particle. 

Since the PSD remains intact, its measurement would single out the DM production model generating distinct velocity distributions. Unfortunately with present observations we can only obtain some estimate of the \emph{coarse grained PSD,} which is the result of averaging of PSD over a range of velocities and over a space region. Naturally, the outcome of the averaging procedure largely depends on the space (angular) and velocity (3-momentum) resolution of the observations. The coarse-grained PSD is known to decrease during the structure formation accompanied by subsequent violent relaxation \cite{Lynden-Bell:1966zjv,STremaine}, when the r.h.s.\ of the Liouville equations above becomes non-zero. Therefore, \emph{the averaging further decreases PSD}, and hence allows one to use the observations to place the lower limits of the true PSD predicted by a DM production mechanism. In particular, considering the maximal value of the coarse grained PSD, one arrives at the inequality  
\begin{equation}
\label{max}
F_{obs}^{max}\leq F_{prod}^{max}\,.
\end{equation}
Alternatively, one can introduce the EMF $D(f)$\,\cite{Kaplinghat:2005sy} as a more detailed proxy of PSD. It is defined as 
\begin{equation}
    \label{def-excess}
D(f)=\int d{\bf x} d{\bf p} \l F({\bf x},{\bf p})-f\r \,\Theta \l F-f\r ,
\end{equation}
where $\Theta$ is the step-function. The function $D(f)$ gives the total excess of the PSD over a given value $f$. Naturally, this function remains constant as far as $F$ does. 

In a realistic situation the EMF also can be used only for the coarse-grained PSD. 
One can check that any mixing (with subsequent coarse-graining) of volumes where $\overline{F} \leq f$ with volumes where $\overline{F} \geq f$ decreases $D(f)$; other mixing (with subsequent coarse-graining) processes leave $D(f)$ unchanged. Hence, for the observed EMF may only become smaller, which means that for any $f$ 
\begin{equation}
\label{max-excess}
D_{obs}(f)\leq D_{prod}(f)\,.
\end{equation}

Below we apply both bounds \eqref{max}, \eqref{max-excess} to place lower limits on sterile neutrino masses in particular models of their production in the early Universe.

\section{Application to a model example: galactic dark matter with local Maxwell distribution and thermal form of the spectrum at production}
\label{sec:analytical}

Since neither DM density profile nor DM spectrum in a galaxy can be directly measured, a set of assumptions are typically introduced \cite{Boyarsky:2008ju,Gorbunov:2008ui} to obtain an estimate of $F^{max}_{obs}$. The well motivated approximation is treating the dSph as a weakly non-equilibrium thermal system, which leads us to the precise form of its PSD as multivariate Gaussian, 
\begin{equation}\label{CG PSD}
    F_{coarse} = \frac{\rho}{(2\pi)^{3/2}m^4\sigma_r\sigma_{\bot}^2} \exp \l -\frac{1}{2} \l \frac{v_r^2}{\sigma_r^2} + \frac{v_\bot^2}{\sigma_{\bot}^2}  \r \r. 
\end{equation}
Here $\rho=\rho(r)$ is the averaged local DM mass density  and subscripts $r$ and $\bot$ refer to the radial and tangential components of the local DM velocity. At any distance the maximal value of PSD is achieved with zero velocities, $v_r=v_\bot=0$, and is proportional only to the quantity 
\[
Q(r)\equiv \frac{\rho(r)}{\sigma_r(r)\sigma_\bot^2(r)}.
\]
The maximum value of \eqref{CG PSD} is reached in the galaxy centre with negligible velocities  
\begin{equation}\label{Maxwell max}
    F_{coarse}^{max} = \frac{1}{(2\pi)^{3/2}m^4} \l \frac{\rho}{\sigma_r\sigma_{\bot}^2} \r _{max} \equiv \frac{Q_{max}}{(2\pi)^{3/2}m^4}.
\end{equation}
Applying the result of the Liouville theorem \eqref{max} we obtain 
\begin{equation}\label{max-PSD-bound}
    \frac{Q_{max}}{(2\pi)^{3/2}\,F_{prod}^{max}} \leq m^4\,. 
\end{equation}
As we mentioned above, EMF also can be calculated only for the coarse-grained PSD. Similarly, we adopt multivariate Gaussian \eqref{CG PSD} as $F_{obs}$ in this case. We also approximate the DM distribution in the galaxy centre with the cored profile, 
\begin{equation}
\label{f-cored}
\rho(r)=\rho_c/(1+r^2/r_c^2)\,, 
\end{equation}
consistent with observations of the dSph. 
Then the  integration in \eqref{def-excess} can be performed analytically over both radial and tangential momenta. The remaining integral over radius
\begin{multline}
    D_{coarse} = \int (4\pi r^2\diff r) \frac{Q}{(2\pi)^{3/2}m^4}\pi^{3/2}2\sqrt{2}m^3\sigma_r^3(1-\beta) \Bigg( \mathrm{Erf}\l \ln^{1/2}\l\frac{Q}{(2\pi)^{3/2}fm^4}\r \r -\\
    - \l\frac{Q}{(2\pi)^{3/2}fm^4}\r^{-1/(1-\beta)} \sqrt{\frac{1-\beta}{\beta}}\mathrm{Erfi}\l \sqrt{\frac{\beta}{1-\beta}}\ln^{1/2}\l\frac{Q}{(2\pi)^{3/2}fm^4}\r\r \Bigg ) \\
    - f\int (4 \pi r^2 dr)\frac {4\pi} {3} 2^{3/2}m^3\sigma_r^3(1-\beta)\ln^{3/2}\l\frac{Q}{(2\pi)^{3/2}fm^4}\r
\label{EMF-obs}
\end{multline}
can be calculated numerically. Here we introduced a new quantity  
\begin{equation}
\label{beta-function}
\beta=\beta(r)\equiv 1 - \sigma_\bot^2(r)/\sigma_r^2(r)
\end{equation}
to characterize the possible velocity anisotropy. 
The integration region is defined by momentum condition $p_r^2=2m^2\sigma_r^2\ln\l\frac{Q}{fm^4}\r \geq 0$. 

We can apply these formulas to the dSphs with parameters estimated in Ref.\,\cite{Munoz:2018}, where the velocities are assumed to be isotropic, i.e.\ $\beta=0$ and dispersions along all three spatial directions are the same and can be related through the one-dimension dispersion $\sigma$ (equal to $\sigma_r$ in this case of spherical symmetry and velocity isotropy).  
In this case, eq.\,\eqref{CG PSD} reads
\begin{equation}\label{CG PSD iso}
    F_{coarse}(r,v)=\frac{\rho(r)}{(2\pi)^{3/2}\sigma^3(r)}\exp \l -\frac{1}{2}\frac{v^2}{\sigma^2(r)} \r.
\end{equation}
The maximal value of PSD is obtained for zero velocities and in the very centre of the galaxy. Denoting the corresponding mass density and velocity dispersion with subscript index $c$ we obtain in this case
\[
 F^{max}_{coarse}=\frac{\rho_c}{(2\pi)^{3/2}\sigma_c^3}\,.
\]

The estimate for the DM density in the centre (actually, the lower
bound on it) can be obtained from the observation
\cite{1986AJ.....92...72R} that, with mild assumptions about
the DM profile one has 
\begin{equation}
  \label{rho0}
  \frac{\rho_c}{M_\odot\;\mathrm{pc}^{-3}} = 148\l \frac{\sigma_c}{\mathrm{km}\;\mathrm{s}^{-1}} \r^2 \l \frac{\mathrm{pc}}{r_h} \r^2,
\end{equation}
where $\sigma_c$ is one-dimension dispersion at the centre. It corresponds to the average radial velocity dispersion in the central part of the galaxy (that is average line-of-sight dispersion) as $\sigma_c^2 \simeq 0.46\bar\sigma^2$. The half-light radius $r_h$ is related to the critical radius from the DM mass density profile \eqref{f-cored} as $r_h \simeq 2r_c$. Therefore, one can obtain for 
$$
Q_{max}=\frac{\rho_c}{\sigma_c^3}
$$
the following numerical estimate through the observable line-of-sight average velocity dispersion in the central part of galaxy
$$
\frac{Q_{max}}{(M_\odot\;\mathrm{pc}^{-3})/(\mathrm{km}\;\mathrm{s}^{-1})^3}= 218\l\frac{\mathrm{km}\;\mathrm{s}^{-1}}{\bar\sigma}\r\l\frac{\text{pc}}{ r_h}\r^2\,.
$$

The EMF integration for isotropic case \eqref{CG PSD iso} leads to:
\begin{multline}
\label{EMF-iso}
    D_{coarse}(f)=\int ((4\pi)^2m^3 r^2\mathrm{d} r) \frac{Q}{(2\pi)^{3/2}m^4} \sigma^3 \Bigg( \sqrt{\frac{\pi}{2}}\mathrm{Erf}\l\ln^{1/2}\l\frac{Q}{(2\pi)^{3/2}fm^4}\r\r - \\
    \sqrt{2}\l\frac{Q}{(2\pi)^{3/2}fm^4}\r^{-1} \ln^{1/2}\l\frac{Q}{(2\pi)^{3/2}fm^4}\r \Bigg)
    - f\int((4\pi)^2 r^2\mathrm{d} r) 2^{3/2}m^3\frac{\sigma^3}{3}\ln^{3/2}\l\frac{Q}{(2\pi)^{3/2}fm^4}\r.
\end{multline}
The integration area is defined by momentum condition $ p^2 = 2m^2\sigma^2\ln\l\frac{Q}{fm^4} \r \geq 0$.

We constrain a simple model of DM production which predicts the thermal form of the sterile neutrino spectra (the Majorana fermion, two degrees of freedom) with normalization parameter ${\cal N}$ tuned to explain the whole DM component, 
\begin{equation}
\label{add-1}
    F_{prod} ={\cal N} \frac{2}{(2\pi)^3}\frac{1}{e^{p/T_{\nu} } + 1}=\frac{11.16}{(2\pi)^3}\frac{\eV}{m}\frac{1}{e^{p/T_{\nu}}+1}\,.
\end{equation}
Here the temperature is normalized to the effective late-time temperature of the relic active neutrinos, related to that of photons as $T_\nu=T_\gamma\,(4/11)^{1/3}$. In particular, this spectrum is very close to what one obtains  for the sterile neutrino production by active neutrino oscillations in lepton-symmetric primordial plasma, the so called Dodelson--Widrow mechanism\,\cite{Dodelson:1993je}. 
The PSD maximum is 
\begin{equation}\label{Fermi max}
    F_{prod}^{max} = \frac{11.16}{2(2\pi)^3}\frac{\text{eV}}{m}\,.
\end{equation}

Using the maximum PSD bound \eqref{max-PSD-bound} one gets
\begin{equation}\label{bound-def}
    m \geq \bar{m} \equiv \l \frac{2Q_{max}(2\pi)^{3/2}}{11.16 \eV} \r^{1/3}.
\end{equation}
The respective 1-$\sigma$ error of $\bar m(Q)$ value can be expressed via errors of the observable quantities, half-light radius and observable average velocity dispersion as:
\begin{equation}\label{bound-err}
    \Delta \bar m = \frac{\bar{m}}{3}\frac{\Delta Q_{max}}{Q_{max}} = \frac{\bar{m}}{3}\sqrt{\l 2\frac{\Delta r_h}{r_h}\r^2 + \l \frac{\Delta\bar\sigma}{\bar\sigma}\r^2}.
\end{equation}
Here we suppose that the two errors are uncorrelated, since the half-light radius is mostly determined from the photometric observations, while the velocity requires a spectrometer.
As the probability density of maximum PSD mass bound is assumed to be normal with mean \eqref{bound-def} and dispersion \eqref{bound-err}, one can obtain one-sided bound $m>m_{95}$ at 95\%CL with the help of error-function as usual,  
\begin{equation}
    0.5 + \int_{\bar{m}}^{m_{95}}\frac{\mathrm{d}m}{\sqrt{2\pi}\Delta m}\exp \l -\frac{1}{2}\frac{(m-\bar{m})^2}{(\Delta m)^2} \r = 0.05\,.
\end{equation}

The EMF corresponding to $F_{prod}^{max}$ is
\begin{equation}\label{Q bound}
\begin{split}
    D_{prod}(f) = 4\pi V_{prim} T^3_{\nu}\\ 
    \times\l \frac{11.16}{(2\pi)^3}\frac{\text{eV}}{m}\int_0^{\ln\l \frac{11.16}{(2\pi)^3}\frac{\eV}{mf} -1\r} \frac{p^2}{\mathrm{e}^{p}+1}\diff p-\frac{1}{3}f\ln^3\l\frac{11.16}{(2\pi)^3}\frac{\eV}{mf}-1\r\r\,,
\end{split}
\end{equation}
where $V_{prim}$ is the initial spatial volume of particles forming the galaxy. Assuming that the total quantity of DM particles $\mathrm{N}_{tot}$ remains constant through the evolution, $V_{prim}$ can be extracted using normalisation condition:
\begin{equation}
    \mathrm{N}_{tot} = \int {d\bf x} {d\bf p}\, F_{prod} = \int {d\bf x}{d\bf p}\,F_{obs} ,
\end{equation}
or in terms of EMF:
\begin{equation}\label{Normal ratio}
    D_{prod}(f=0) = D_{obs}(f=0).
\end{equation}

In practice, the inequality \eqref{max-excess} can be considered only for $f\geq f_{min}=\frac{Q_{min}}{(2\pi)^{3/2}m^4}$ and then, one can obtain the equation for mass bound value $\bar m$

\begin{equation}\label{EMF-bound}
    D_{coarse}\left(f=\frac{Q_{min}}{ (2\pi)^{3/2} \bar{m}^4 }\right) - D_{prod}\left(f=\frac{Q_{min}}{(2\pi)^{3/2}\bar{m}^4}\right) = 0.
\end{equation}

For isotropic case, the minimum value $Q_{min}$ we assume to be implemented at $r\simeq10r_c$.
The corresponding errors following from \eqref{EMF-bound} and one-sided bounds at 95\%CL can be calculated using Monte--Carlo method for the numerical integration. 

The lower limits on the sterile neutrino mass obtained within the two approaches for the set of galaxies are presented in Tab.\,\ref{tab:dSph}. 
\begin{table}[!htb]
\tbl{Lower limits on the sterile neutrino DM mass (non-resonant production mechanism) obtained with application of the EMF and application of maximum PSD and the corresponding one sigma error bars ($\pm \Delta m$) obtained for the set of dSph galaxies, whose mass density and velocity dispersions in the galaxy central  parts (and their uncertainties) are given in Ref.\,\cite{Munoz:2018}. All values are in keV.}
    {\centering
    \resizebox{\columnwidth}{!}{
    \begin{tabular}{|l|c|c|c|c|c|c|}
 \hline
 \multicolumn{1}{|c|}{Method $\rightarrow$} & \multicolumn{3}{|c|}{EMF}  & \multicolumn{3}{|c|}{Maximum PSD} \\ \hline
Object $\downarrow$ & $m$ & $+\Delta m $ & $-\Delta m$ & $m$ & $+\Delta m$ & $-\Delta m$ \\\hline
        Sculptor & 5.34 & 0.26 & 0.25 & 1.90 & 0.10 & 0.10 \\ \hline
        Fornax & 2.36 & 0.06 & 0.06 & 0.84 & 0.02 & 0.02 \\ \hline
        Carina & 5.59 & 0.39 & 0.30 & 1.99 & 0.12 & 0.12 \\ \hline
        NGC 2419 & 39.00 & 1.64 & 1.47 & 13.91 & 0.53 & 0.53 \\ \hline
        UMa II & 10.00 & 0.88 & 0.68 & 3.55 & 0.26 & 0.26 \\ \hline
        Leo T & 8.56 & 0.96 & 0.76 & 3.04 & 0.3 & 0.3 \\ \hline
        Segue 1 & 36.34 & 4.49 & 3.30 & 12.92 & 1.33 & 1.33 \\ \hline
        Leo I & 5.43 & 0.33 & 0.23 & 1.94 & 0.10 & 0.10 \\ \hline
        Sextans & 4.32 & 0.26 & 0.22 & 1.54 & 0.08 & 0.08 \\ \hline
        UMa I & 6.39 & 0.38 & 0.30 & 2.28 & 0.12 & 0.12 \\ \hline
        Willman 1 & 32.11 & 4.59 & 3.08 & 11.43 & 2.14 & 1.33 \\ \hline
        Leo II & 8.27 & 0.31 & 0.30 & 2.94 & 0.11 & 0.11 \\ \hline
        Leo V & 22.43 & 8.53 & 4.70 & 7.92 & 2.36 & 1.96 \\ \hline
        Leo IV & 13.70 & 3.78 & 1.9 & 4.86 & 0.90 & 0.90 \\ \hline
        ComBer & 16.61 & 1.12 & 1.09 & 5.91 & 0.40 & 0.40 \\ \hline
        CVn II & 16.89 & 2.55 & 1.95 & 5.98 & 0.77 & 0.77 \\ \hline
        CVn I & 4.11 & 0.11 & 0.09 & 1.47 & 0.04 & 0.04 \\ \hline
        Bootes II & 18.83 & 6.06 & 3.28 & 6.79 & 1.70 & 1.70 \\ \hline
        Bootes I & 10.70 & 0.99 & 0.68 & 7.60 & 0.96 & 0.54 \\ \hline
        Munoz 1 & 224.89 & 63.79 & 41.42 & 78.36 & 17.14 & 17.14 \\ \hline
        UMi & 4.11 & 0.18 & 0.16 & 1.46 & 0.06 & 0.06 \\ \hline
        Hercules & 8.64 & 0.90 & 0.76 & 3.06 & 0.30 & 0.30 \\ \hline
        Draco & 6.38 & 0.29 & 0.27 & 2.28 & 0.10 & 0.10 \\ \hline
        NGC 7492 & 94.50 & 32.96 & 15.62 & 34.99 & 9.73 & 9.73 \\ \hline
        Eridanus II & 7.42 & 0.66 & 0.51& 2.65 & 0.23 & 0.20 \\ \hline 
     \end{tabular}}
    \label{tab:dSph}}
\end{table}
First, one observes that EMF is generically more restrictive, that maximum PSD. Second, 
10 out of these 25 galaxies reveal $m>10$\,keV, 5 give $m>20$\,keV and 3 suggest even $m>30$\,keV. These are  extremely serious limits with potential to close many models of sterile neutrino DM production and similar candidates which form WDM. Third,   
most limits from the maximal PSD given here are numerically much stronger than those in Ref.\,\cite{Gorbunov:2008ui}, which may be attributed to new observational data and astronomical analysis (compare the results for Coma Berenices, Leo IV and
Canes Venaciti II). 

However, the limits obtained using this approach in Tab.\ref{tab:dSph} vary broadly between different galaxies. The four cases of NGC 2419, Segue 1, Willman 1, Munoz 1 and NGC 7492 look extremely suspicious, providing limits which are stronger by almost two orders of magnitude (as compared to all the rest) even including the error bars, which we evaluated by making use of the uncertainties on average velocities, critical radii and matter densities presented in Ref.\,\cite{Munoz:2018}, and applying eq.\,\eqref{bound-err}. This result asks for a more careful analysis of the observational data, which we perform below with the help of GravSphere numerical code aimed to reconstruct the DM dynamics from observations of galaxy star velocities.

\section{Reconstruction of galaxy dark matter phase space density from numerical fit to observational data on the galaxy stars}
\label{sec:numerical}

One can try to reconstruct the coarse-grained PSD using the observed data. Namely, we can use observations of the Doppler effects in the galaxy stars, which gives the average dispersion of the star velocity projected on the line of sight $\sigma_{LOS*}(r)$, and photometric observations of average surface star density $\Sigma_*(r)$ to estimate the functions entering the PSD \eqref{CG PSD}.     

Since both stars and DM particles are nonrelativistic inside a galaxy, their radial matter 
density distributions, $\rho_*$ and $\rho$, and corresponding radial $\sigma_r^2$ and tangential $\sigma_\bot^2\equiv(1-\beta)\sigma_r^2$ velocity dispersions (star $*$ subscript hereafter refers to the stellar component) satisfy the Radial Jeans Equations 
(RJEs) 
\begin{eqnarray}\label{RJE stellar}
    \frac{1}{\rho_*}\frac{\partial}{\partial r}(\rho_*\sigma_{r*}^2) + 2\frac{\beta_*\sigma_{r*}^2}{r} = -\frac{GM_{tot}(r)}{r^2},\\
\label{RJE DM}
    \frac{1}{\rho}\frac{\partial}{\partial r}(\rho\sigma_r^2) + 2\frac{\beta\sigma_r^2}{r} = -\frac{GM_{tot}(r)}{r^2},
\end{eqnarray}
where $M_{tot}$ is the total mass inside the radius $r$, which is saturated by the DM component.

The stellar component PSD for a particular galaxy can be reconstructed with the help of non-parametric Jeans code GravSphere \cite{GravSphere_ref}, which solves RJE for the stellar component \eqref{RJE stellar} aiming to obtain a good fit to the corresponding observational stellar data sets, 
$\sigma_{LOS*}$ and $\Sigma_*$.

The solutions to both RJE for DM and stellar components \eqref{RJE stellar}, \eqref{RJE DM} can be written as follows
\begin{eqnarray}
  \label{sigma-stellar-sol}
    \sigma_{r*}^2(r) &=& \frac{1}{g_*(r)\rho_*(r)}\int_r^\infty \frac{1}{s^2}GM_{tot}(s)g_*(s)\rho_*(s)\mathrm{d}s, \\
  \label{sigma-DM-sol}
    \sigma_r^2(r) &=& \frac{1}{g(r)\rho(r)}\int_r^\infty \frac{1}{s^2}GM_{tot}(s)g(s)\rho(s)\mathrm{d}s,
\end{eqnarray}
where $g_{(*)}(r)\equiv\exp \l 2\int\frac{\beta_{(*)}(r)}{r}\mathrm{d}r \r,$ and boundary conditions for DM and stellar components, respectively, are $\rho\sigma_r^2(r\rightarrow\infty)=0$ and $\rho_*\sigma_{r*}^2(r\rightarrow\infty)=0$. In case of dSphs the total mass is dominated by the DM component, 
\begin{equation}
    \label{total-mass}
M_{tot}(r)=4\pi\int_0^r \rho(s)s^2{\text d}s\,. 
\end{equation}
For a given DM profile $\rho(r),$ (we utilize the profile parameterization advocated in Refs.\,\cite{Read_2018,Collins_2021}) star velocity dispersion \eqref{sigma-stellar-sol}, matter distribution \eqref{total-mass}, and stellar velocity anisotropy function $\beta(r)$ (see below),
the GravSphere code calculates following \eqref{sigma-stellar-sol} the star surface mass density $\Sigma_*(r)$ and line-of-sight velocity dispersion 
\begin{equation}
     \sigma_{LOS}^2(R) = \frac{2}{\Sigma_*(R)}\int_R^\infty \l 1-\beta_*(r)\frac{R^2}{r^2}\r\frac{\rho_*(r)\sigma_{r*}(r)r}{\sqrt{r^2-R^2}}\mathrm{d}r\,.
\end{equation}
The latter two quantities are compared with the observational data sets 
and the stellar and DM object mass, parameters of anisotropy $\beta_*(r)$ are adjusted to obtain a good fit to the data.
The fitting is implemented through {\sc emcee} affine invariant Markov Chain Monte Carlo (MCMC) sampler \cite{MCMC_ref}. For each galaxy the code runs a chain with 250 walkers  with $5\times 10^4$ steps each, discarding
first 75\% of the steps as a 'burn-in'. The remaining $3.125\times10^6$ samples thus provide an approximation of distribution of the required stellar and matter densities and stellar velocity dispersion. Their statistical characteristics reflect those of the data: e.g.\ at a given radius more models exhibit the most favourable by the data stellar and matter densities, the models with worth fit to the data are rare. This feature justifies using these distributions of the DM density and velocity dispersion in the formulas limiting the DM mass and getting the statistical limits at a given confidence level. In practice, the final confidence intervals for the parameters of interest are obtained on a smaller subset (1000 samples) of the whole MCMC output.

The original version of GravSphere code does not calculate the radial velocity dispersion for the DM component, so we are using \eqref{sigma-DM-sol} to obtain the DM velocity dispersion needed for the numerical estimates of the PSD distributions, which are used to place the lower limits on sterile neutrino mass.

In our analysis we selected 20 galaxies from the dSph catalogue  \cite{2012AJ....144....4M} (with the latest 2021 update given at the link\,\footnote{\url{https://www.cadc-ccda.hia-iha.nrc-cnrc.gc.ca/en/community/nearby/}}),
which have sufficient numbers of individually observed stars available in the SIMBAD database\,\footnote{\url{https://simbad.cds.unistra.fr/simbad/}}.
There are two sets of star observations: \emph{photometric,} where only the star position and brightness are measured, and \emph{kinematic,} where the Doppler effect is measured as well. The first set is used to determine the stellar density, while the second set is also used to estimate the stellar velocity distribution. Each set is subjected to the binning procedure
\cite{GravSphere_ref, Read_2018, Genina_2020, Collins_2021} along the radial coordinate.
The size of each bin is individually chosen to include the same number of stars in each bin. So, the solution of the RJEs for each galaxy must fit to the data presented as two sets of binned data $\sigma_{LOS*}[i]$ and $\Sigma_*[i]$. The latter is defined as the surface star number density, that is the number of stars in the bin divided by the bin size and average (for this subsample) radius, i.e.,  distance to the galaxy centre. The former is the velocity dispersion along the line-of-sight calculated from the given line-of-sight velocities of stars in the bin. The optimal number of measurements (star positions or line-of-sight-velocities) per bin was chosen as $N_{\text{bin}} = \lfloor \sqrt{N_{\text{photometric}}} \rfloor$ and 
$N_{\text{binkin}} = \lfloor \sqrt{N_{\text{kinematic}}} \rfloor$ 
for each galaxy. We limited ourselves only to galaxies which would provide not less than 6 bins (both photometric and kinematic), which limits the amount of viable galaxies, with number of kinematic measurements usually being more constraining. The entire set of the selected 20 galaxies with relevant astronomical parameters and references are presented in Tab.\,\ref{tab:galaxies-dataset}.
\begin{table}[!htb]
\tbl{Set of 20 dSphs adopted for our PSD analysis.}
    {\centering
    \resizebox{\columnwidth}{!}{%
    \begin{tabular}{|l|c|c|c|c|c|}
    \hline
        \hfil Name & $D$ & $m$ & $N$& $N$  & References \\ 
        \hfil in SIMBAD database &  [kpc] & [Vega mag] & (photometric) & (kinematic) & \\ \hline
        Andromeda V & 810 & 15.3 & 94 & 94 &  \cite{2018MNRAS.479.4136K},	\cite{2012AJ....144....4M}
 \\ \hline
        Aquarius Dwarf & 940 & 14.8 & 176 & 70 &  \cite{2011ApJS..192....6L},	\cite{2014MNRAS.445..881C}
 \\ \hline
        Bootes Dwarf Spheroidal Galaxy & 66 & 12.8 & 88 & 55 &  \cite{2020ApJ...893...47D}, \cite{2012AJ....144....4M}
 \\ \hline
        Carina dSph & 105 & 11.0 & 1118 & 729 &  \cite{2020ApJ...893...47D}, \cite{2012AJ....144....4M} \\ \hline
        Cetus Dwarf Galaxy & 790 & 13.2 & 781 & 116 &  \cite{2018MNRAS.479.4136K},	\cite{2012AJ....144....4M}
 \\ \hline
        Coma Dwarf Galaxy & 44 & 14.1 & 80 & 51 & \cite{2020ApJ...893...47D}, \cite{2012AJ....144....4M} \\ \hline
        CVn I dSph & 218 & 13.1 & 217 & 91 & \cite{2020ApJ...893...47D}, \cite{2012AJ....144....4M} \\ \hline
        Dra dSph & 76 & 10.6 & 647 & 205 & \cite{2020ApJ...893...47D}, \cite{2012AJ....144....4M} \\ \hline
        Fornax Dwarf Spheroidal & 147 & 7.4 & 3899 & 3207 &  \cite{2020ApJ...893...47D}, \cite{2012AJ....144....4M} \\ \hline
        Hercules Dwarf Galaxy & 132 & 14.0 & 64 & 34 &  \cite{2020ApJ...893...47D}, \cite{2012AJ....144....4M} \\ \hline
        Leo A & 800 & 13.3 & 234 & 64 & \cite{2018ApJ...861...49H}, \cite{2014MNRAS.445..881C}
 \\ \hline
        NGC 6822 & 500 & 8.1 & 891 & 306  & \cite{2011ApJS..192....6L}, \cite{2012AJ....144....4M}
 \\ \hline
        PegDIG & 900 & 12.5 & 109 & 106  & \cite{2018ApJ...861...49H},	\cite{2014MNRAS.445..881C}
 \\ \hline
        Sculptor Dwarf Galaxy & 84 & 8.6 & 1839 & 1073 & \cite{2020ApJ...893...47D}, \cite{2012AJ....144....4M} \\ \hline
        Sextans dSph & 86 & 10.4 & 505 & 218  & \cite{2020ApJ...893...47D}, \cite{2012AJ....144....4M} \\ \hline
        Sgr dIG & 1040 & 13.6 & 77 & 45 & \cite{2011ApJS..192....6L}, \cite{2012AJ....144....4M}
 \\ \hline
        UMi Galaxy & 76 & 10.6 & 226 & 103  & \cite{2020ApJ...893...47D}, \cite{2012AJ....144....4M} \\ \hline
        WLM Galaxy & 920 & 11.1 & 237 & 82  & \cite{2011ApJS..192....6L},	\cite{2014MNRAS.445..881C}
 \\ \hline
        Z 126-111 & 233 & 12.0 & 389 & 261 & \cite{2020ApJ...893...47D}, \cite{2012AJ....144....4M} \\ \hline
        Z 64-73 & 300 & 10.0 & 1312 & 415  & \cite{2018ApJ...861...49H}, \cite{2012AJ....144....4M}
 \\ \hline
    \end{tabular}%
    }
    \label{tab:galaxies-dataset}
}
\end{table}
Likewise, all individual stars used for the analysis, their relevant characteristics and corresponding references may be found at the link \cite{Koreshk_github}.   

A typical dSph from the chosen dataset has very low luminosity as we confirm from our analysis, since its total mass inferred from solving RJE much exceeds its stellar mass, 
\begin{equation}
    M_* \ll M_{tot} \implies M_{DM} \simeq M_{tot}.
\end{equation}
Hence, the value of $M_*$ does not affect any calculations.  However, we take the stellar mass to have the same parametric form, as implemented in GravSphere code, and the total stellar mass to depend on the apparent magnitude $m$ (in Vega magnitudes) as follows:
\begin{equation}
    M_* = 10^{-m/2.5}\l \frac{D}{D_{Vega}} \r^2 M_{Vega}.
\end{equation}

Finally, to solve numerically eqs.\,\eqref{RJE stellar} and \eqref{RJE DM}  the velocity anisotropies $\beta_*(r)$ and $\beta(r)$ must be specified. The former can be probed with the stellar observations while latter cannot be measured either directly or indirectly. GravSphere adopts the following parametrization for the stellar velocity anisotropy 
\begin{equation}
\label{fit-anis}
\beta=\beta_0+\frac{\beta_\infty-\beta_0}{1+(r/r_0)^n}\,,
\end{equation}
where four parameters $\beta_0$, $\beta_\infty$, $r_0$ and $n$ are determined for each galaxy from the fit to the stellar observational data. 
As we consider the early-time evolution of the visible and dark components generally different, there is no reason to consider stellar and DM velocity anisotropies to be equal.
In our study we follow Ref.\,\cite{Alvey:2020xsk}, and use the same parametric form \eqref{fit-anis} for both stellar and DM components.  While the parameters of the stellar component for each galaxy are constrained from the fit to the observational data, the parameters of the DM component are randomly chosen for each of the MC models involved in calculating the DM velocity dispersion. Following Ref.\,\cite{Alvey:2020xsk} we assume the DM velocity anisotropy parameters to be uniformly distributed within segments as $\beta_{\infty}\in[0;0.56]$, $r_0\in[0;3.10]$, $n\in[0.73;1.36]$ and $\beta_0$ is fixed at zero.

\begin{figure}[!htb]
    \centerline{
    \includegraphics[width=0.33\columnwidth]{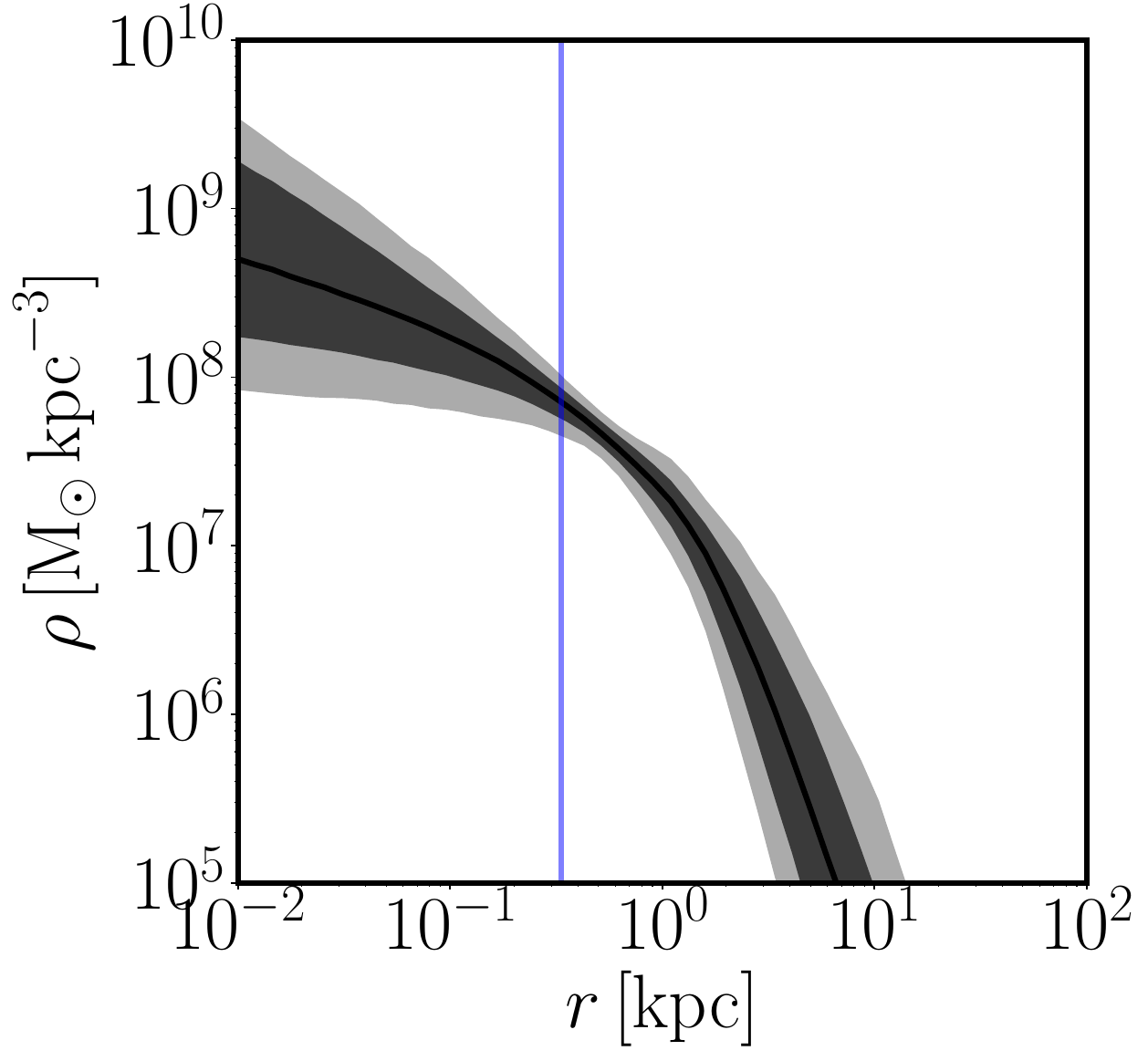}
    \includegraphics[width=0.33\columnwidth]{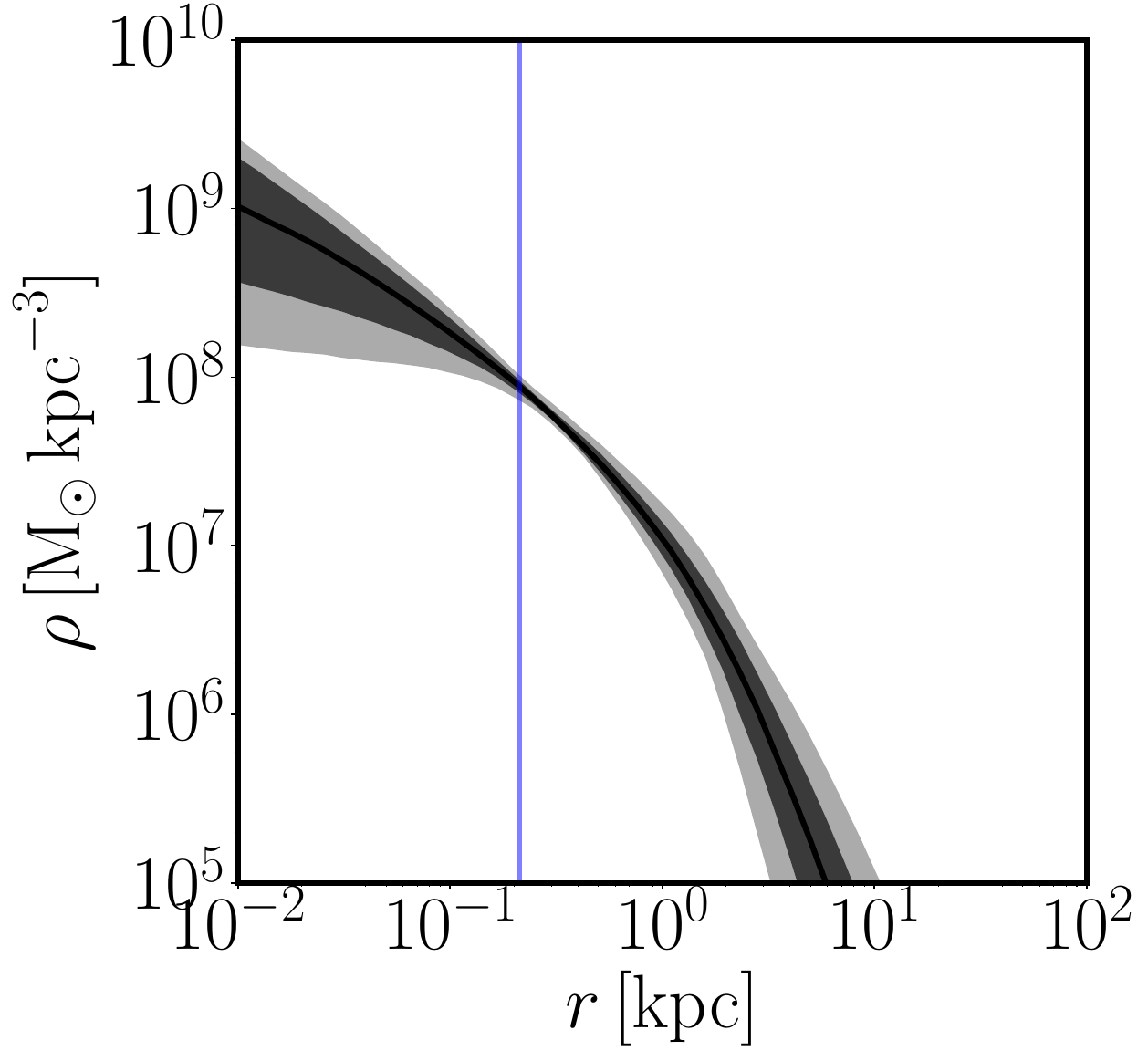}
    \includegraphics[width=0.33\columnwidth]{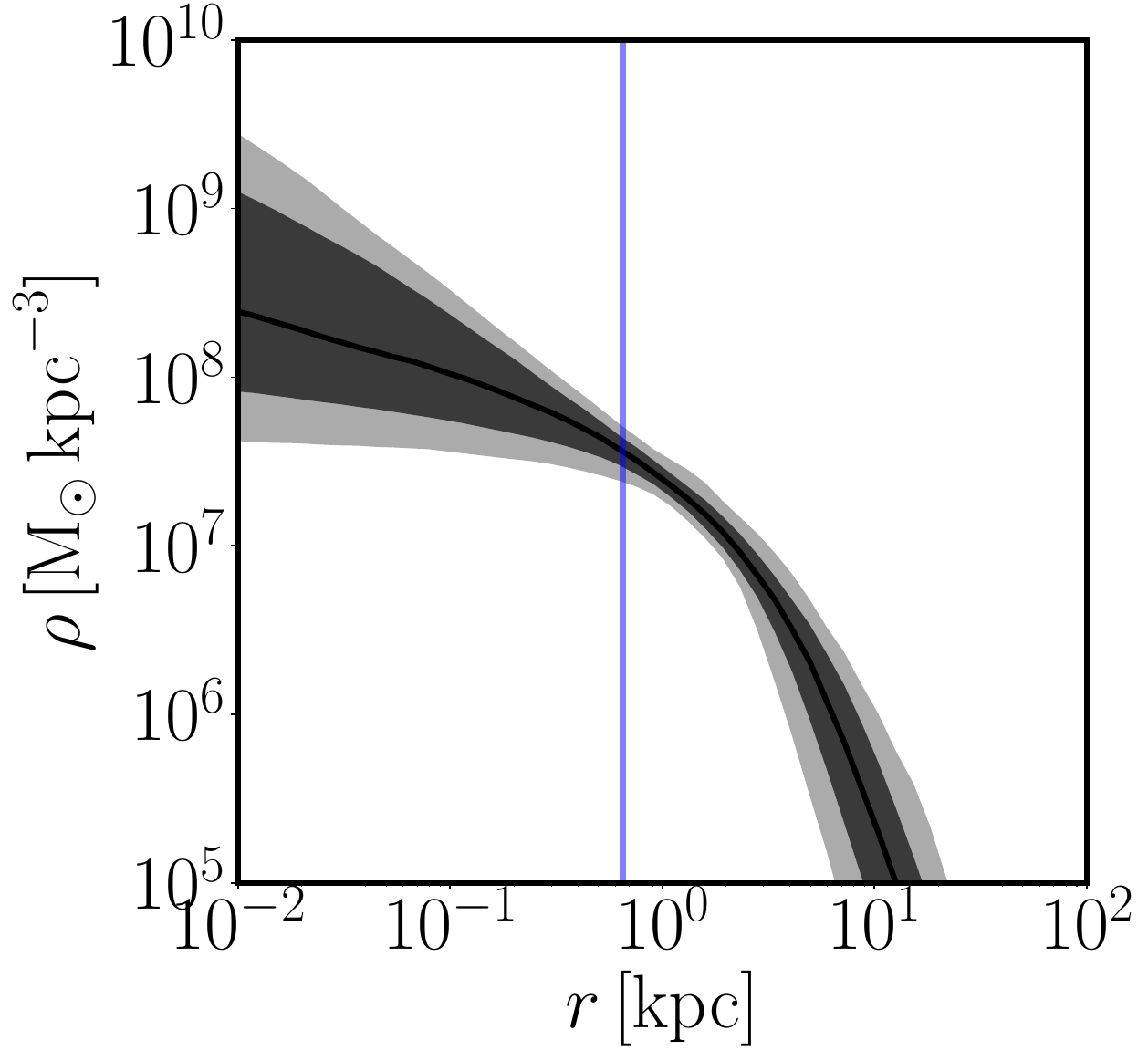}}
    \centerline{
    \includegraphics[width=0.33\columnwidth]{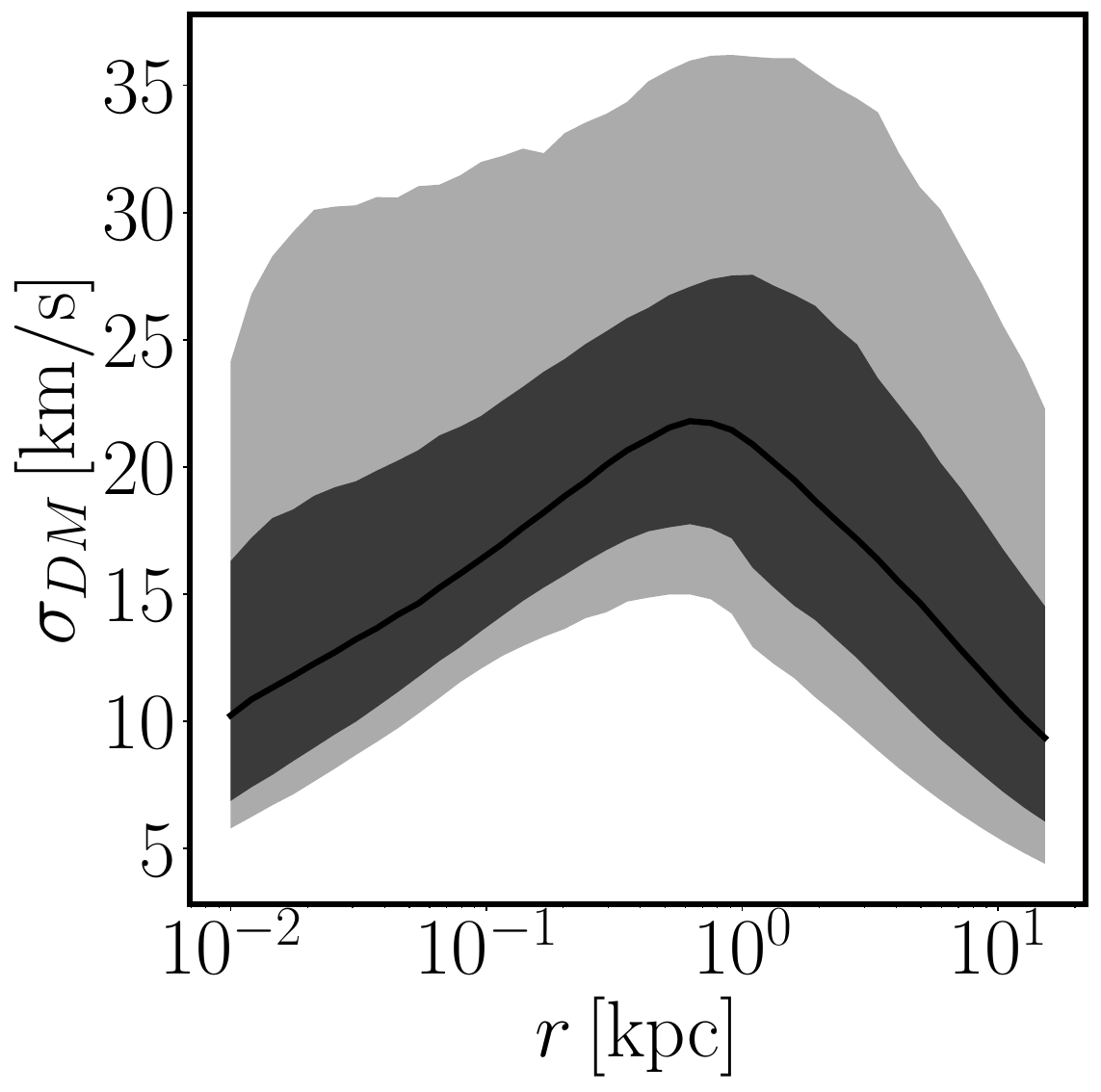}
    \includegraphics[width=0.33\columnwidth]{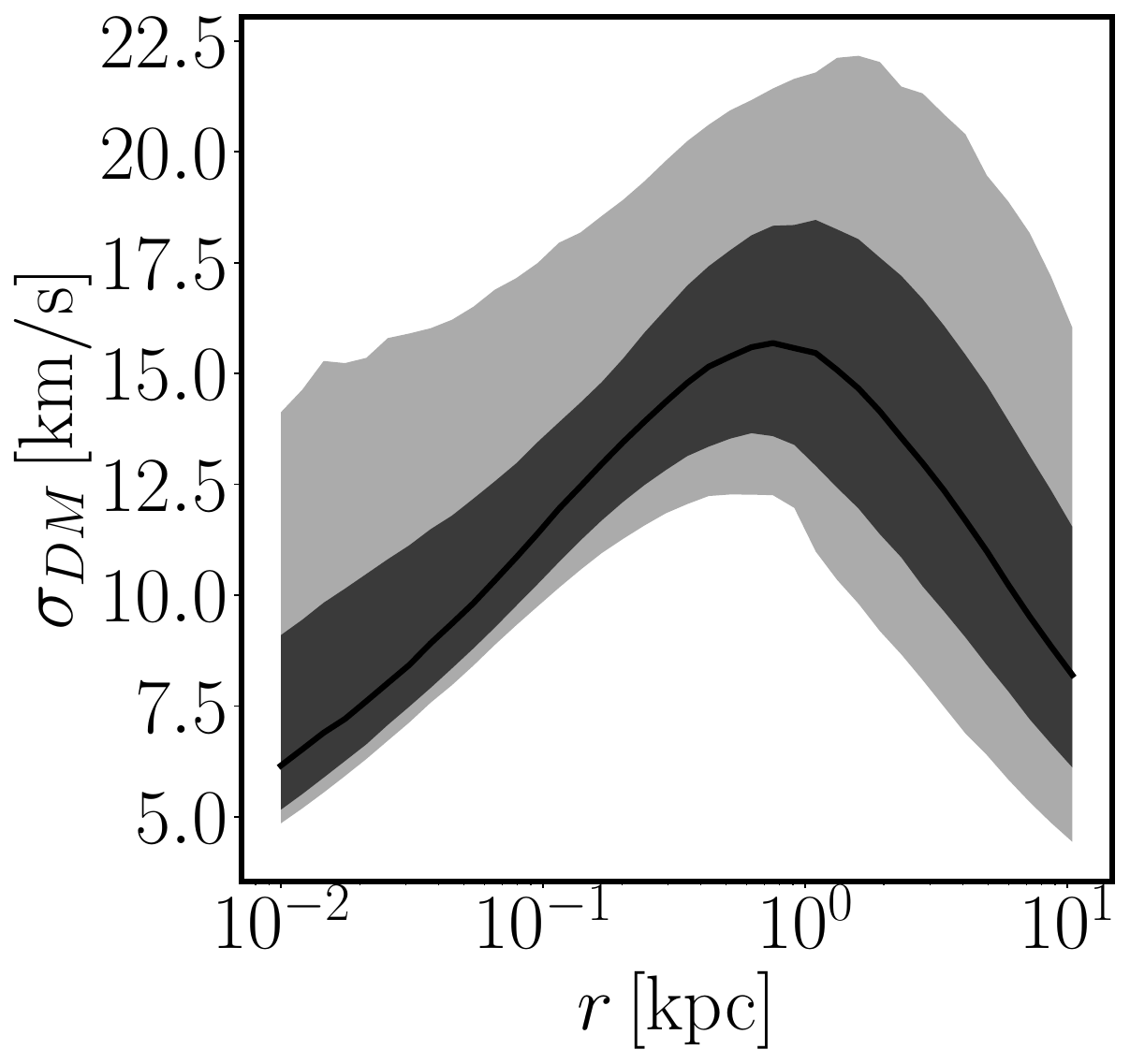}
    \includegraphics[width=0.33\columnwidth]{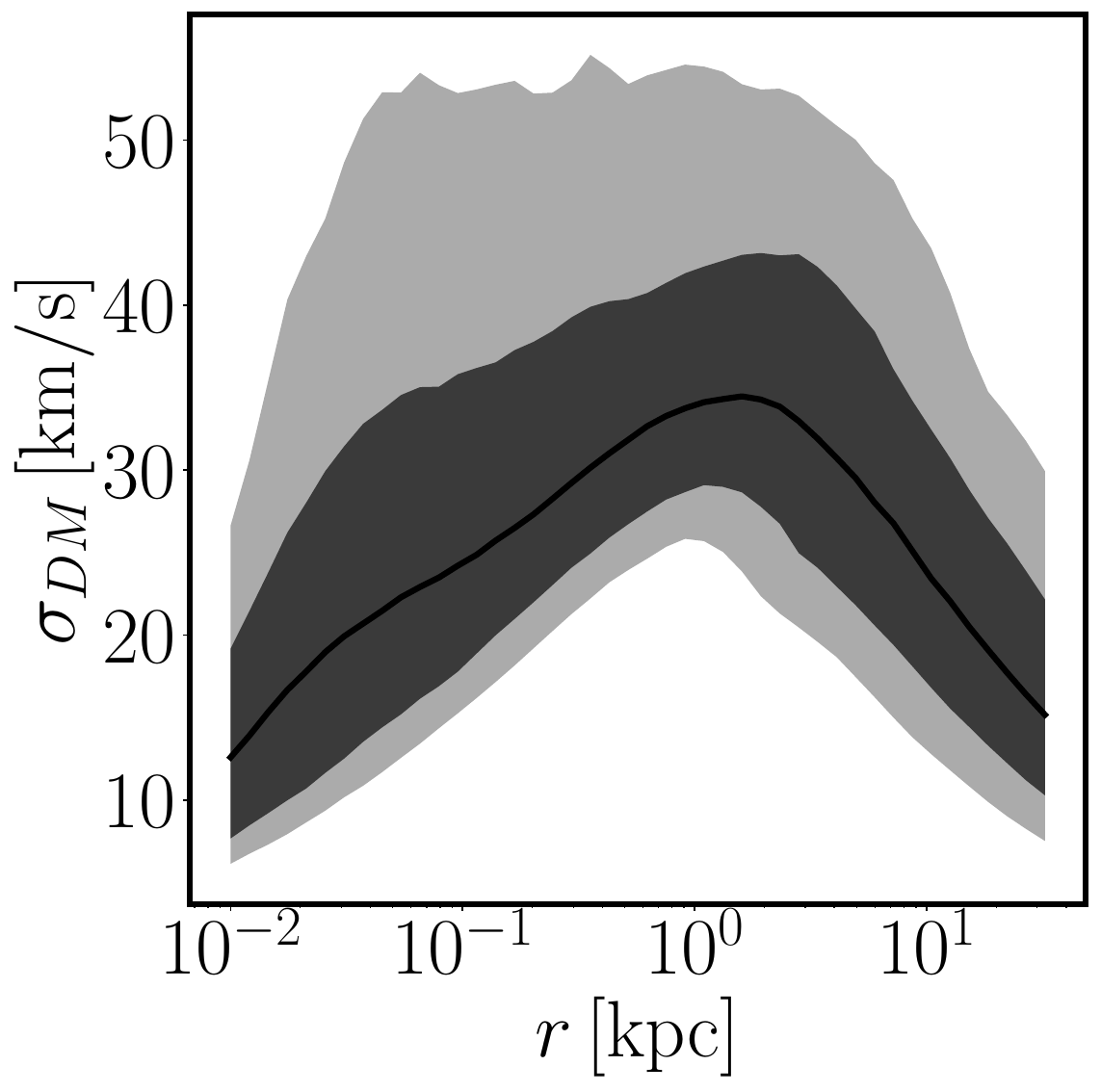}}
    \caption{GravSphere numerical approximation to functions $\rho(r)$ (top) and $\sigma_r(r)$ (bottom) for Aquarius Dwarf (left panel), Sculptor Dwarf Galaxy (middle panel) and WLM Galaxy (right panel). 
    Solid lines refer to the central values, the 65\% CL and 95\% CL regions are shaded with dark and light colours. The dark blue vertical lines indicate the position of half-light radius $r_h$. 
    }
    \label{fig:Gravsphere-outputs}
\end{figure}

With all these considerations our procedure is straightforward. For each galaxy in the sample we use the observational data and solve the Jeans equations \eqref{RJE stellar} and \eqref{RJE DM} with routine GravSphere extended to solve the RJE for DM as we described above. It yields a sample of numerical approximations to the functions $\rho(r)$, $\sigma_r(r)$, $\beta_*(r)$ and similar for DM; the latter determines $\sigma_\bot(r)$ via \eqref{beta-function}. From the sample of numerical solutions we select the best fit to the observational data and corresponding 68\% and 95\% confidence regions. For three typical galaxies the results are presented in Fig.\,\ref{fig:Gravsphere-outputs}, the remaining results can be found at the link \cite{Koreshk_github}.   
The profiles we obtain in this way typically exhibit a mild cusp in the centre, at $r\to 0$, but the lower edges of the 2-sigma contours, see upper plots of Fig.\,\ref{fig:Gravsphere-outputs}, are almost horizontal lines, and hence close to the cored profile \eqref{f-cored} we used in the traditional analysis of the DM phase space density in Sec.\,\ref{sec:analytical}. 
Then we substitute these estimates into eq.\,\eqref{CG PSD} and obtain the radial dependent coarse grained PSD. 

\begin{figure}[!htb]
    \centerline{
    \includegraphics[width=0.33\columnwidth]{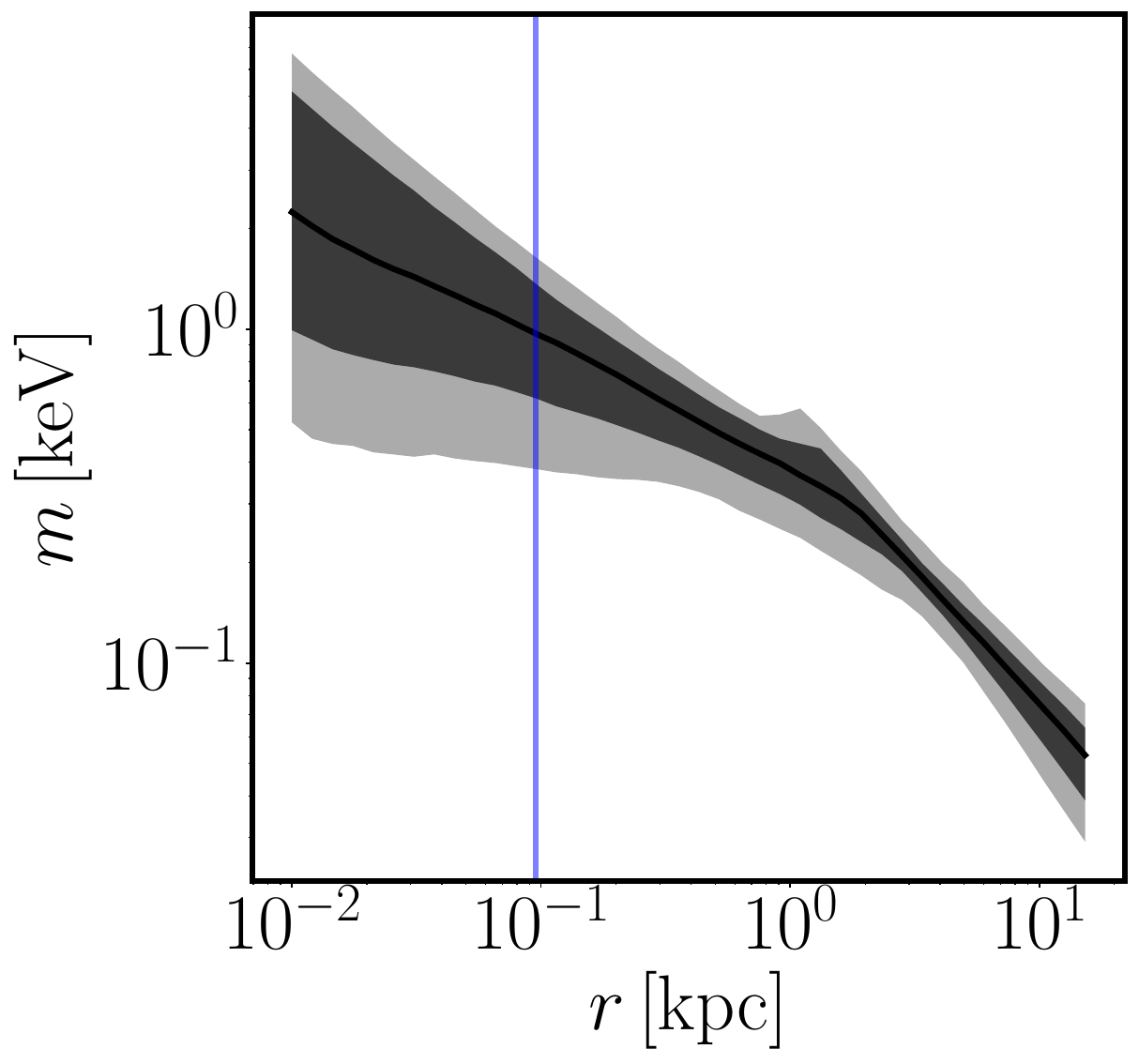}
    \includegraphics[width=0.33\columnwidth]{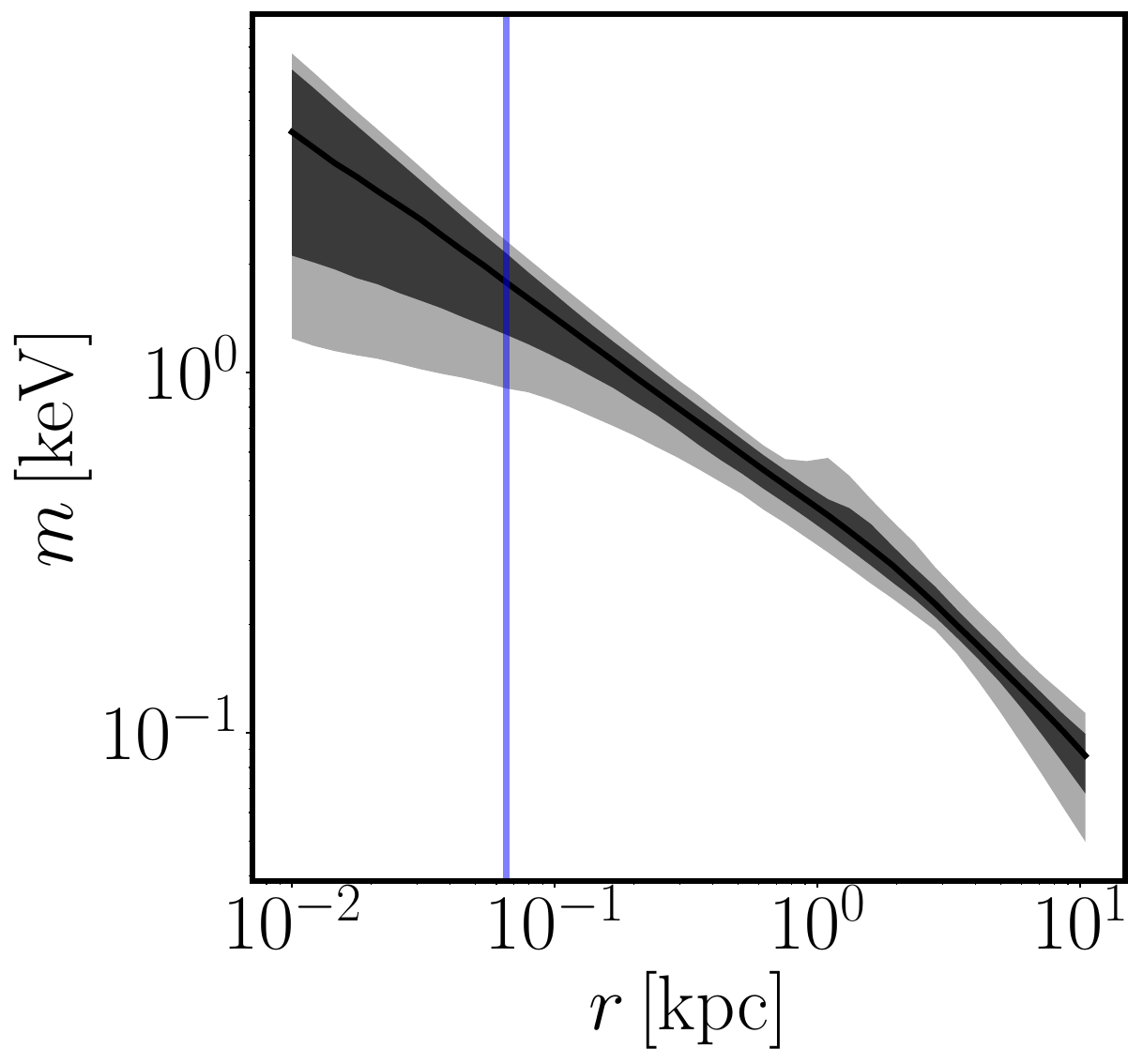}
    \includegraphics[width=0.33\columnwidth]{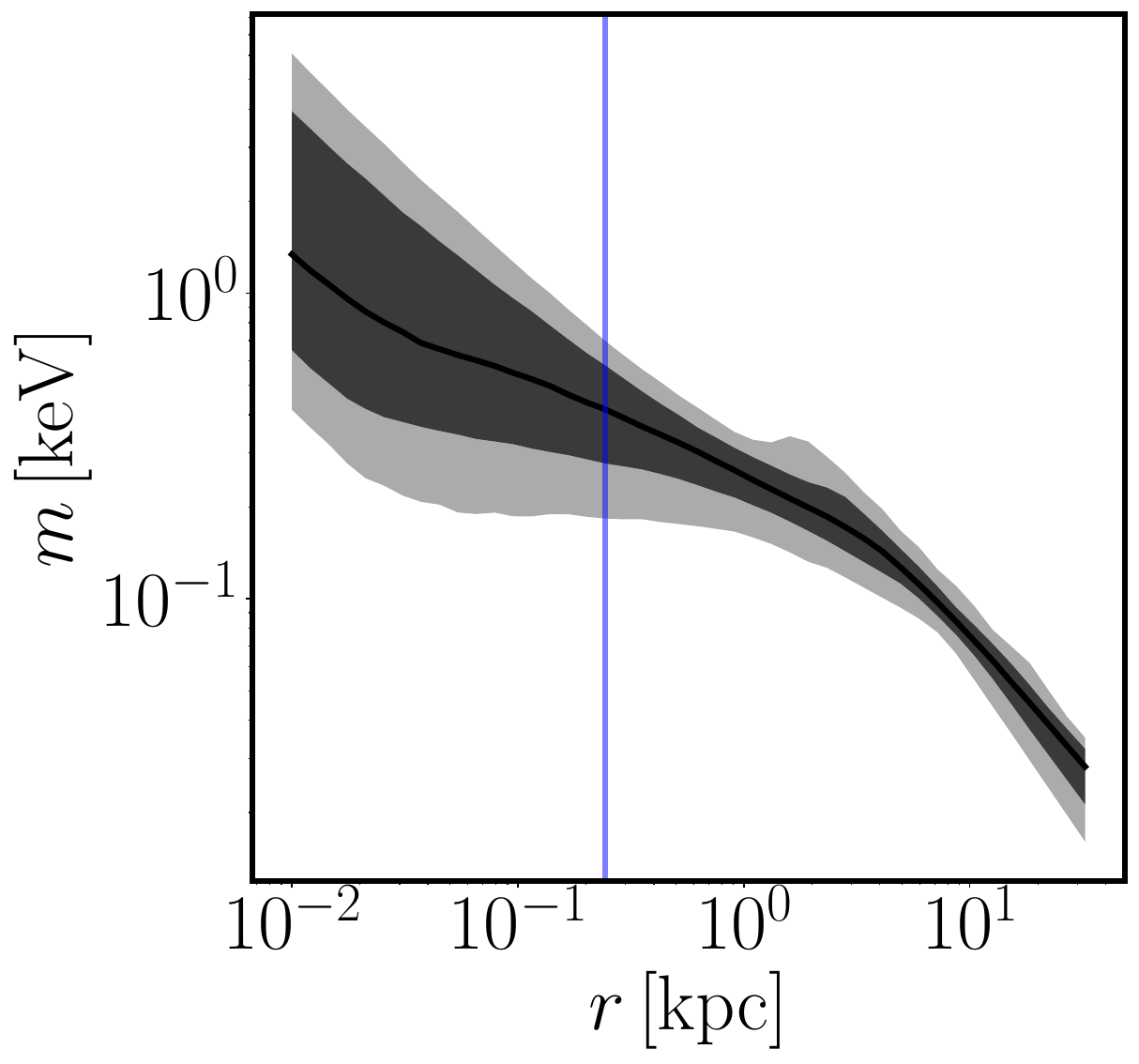}}
    \caption{Radius-dependent best fit values (solid lines), 1- and 2-$\sigma$ regions of the sterile neutrino mass obtained from numerical fit to the Aquarius Dwarf data (left), Sculptor Dwarf Galaxy (middle) and WLM Galaxy (right), see the main text for details. The blue vertical lines indicate the radius we use to place the lower bound on the sterile neutrino DM mass.}
    \label{fig:m-r-examples}
\end{figure}

To use the maximal PSD estimator we analyse $F_{coarse}(r)$ with $v_r=v_\bot=0$ to find its both best fit values and the corresponding confidence intervals at each radius. It can be compared to the predictions of the DM production mechanism in the early Universe via \eqref{max}, where for the latter we use the estimate \eqref{Fermi max}. In Fig.\,\ref{fig:m-r-examples} 
we present the typical estimates for the possible exclusion masses with PSD taken at different radius. More plots for other galaxies are present at the link \cite{Koreshk_github}.
One concludes that the strongest limits come from the central part of the galaxies, naively from PSD at the very centre, $r\to0$. However, in this region typically there are no stellar observational data, and hence the numerical solution we obtain is rather uncertain.  

To get the reliable bounds we prefer to use the solution in the region with sufficient number of observed stars. Hence we adopt the following procedure to fix the radius $\bar r$ at which place the bound. For each galaxy we take among all $N$ observed stars the star at the maximal distance from the centre and call this distance  $r_{max}$. Then we define a set of equal bins in radius with size equal to $r_{max}/\sqrt{N}$. 
We count the number of stars in each bin, chose the bin with maximal population and find the bin with smaller radius where the population is approximately two times smaller. The corresponding radius $\bar r$ we take as the minimal radius at which our numerical solution is still sufficiently supported by the observational data. We chose this radius to place the lower limits on the sterile neutrino DM mass.   

To use the EMF estimator we repeat the same procedure but adopt eq.\,\eqref{EMF-obs} and \eqref{Q bound} instead of \eqref{max} and \eqref{Fermi max}. 

The numerical results are presented in Tab.\,\ref{tab:main-results} for all 20 galaxies from our chosen set. 
\begin{table}[!htb]
\tbl{Main results of numerical analysis in case of non-resonantly produced sterile neutrino DM. For each dSph we show the radius $\bar r$, where the PSD is taken being calculated with the help of GravSphere routine and the original procedure described in the main text. Then for both methods we present the central value of the sterile neutrino mass $\bar m$, with our formulas it provides the best fit to the observations. The columns ``$m>..$'' contain the inferred lower limits at 95\% CL. All masses are in keV.}
{\centering
    \resizebox{\columnwidth}{!}{%
    \begin{tabular}{|l|c|c|c|c|c|}
    \hline
 \multicolumn{1}{|c}{ Method $\rightarrow$ } & 
  \multicolumn{1}{c|}{} & 
 \multicolumn{2}{|c|}{Maximum PSD}  
 & \multicolumn{2}{|c|}{EMF}  
 \\ \hline
Object $\downarrow$ & $\bar r$, pc & $\bar m$ & $m>..$ & $\bar m$ & $m>..$ \\\hline
     Andromeda V & 115 & 1.23 & 0.71 & 2.62 & 1.45\\ \hline
        Aquarius Dwarf & 79 & 1.04 &  0.46  & 2.14 & 0.88\\ \hline
        Bootes Dwarf Spheroidal Galaxy & 79 & 1.27 & 0.82 & 2.53 & 1.57\\ \hline
        Carina dSph & 54 & 1.21 & 0.48 & 2.38 & 0.85\\ \hline
        Cetus Dwarf Galaxy & 167 & 1.35 & 0.82 & 2.87 & 1.68\\ \hline
        Coma Dwarf Galaxy & 625 & 0.61 & 0.55 & 1.09 & 0.98\\ \hline
        CVn I dSph & 65 & 0.98 & 0.41 & 2.03 & 0.78\\ \hline
        Dra dSph & 65 & 0.98 & 0.47 & 1.88 & 0.83\\ \hline
        Fornax Dwarf Spheroidal & 115 & 0.81 & 0.47 & 1.67 & 0.93\\ \hline
        Hercules Dwarf Galaxy & 44 & 3.17 & 1.23 & 6.91 & 2.49\\ \hline
        Leo A & 138 & 1.28 & 0.77 & 2.57 & 1.47\\ \hline
        NGC 6822 & 244 & 0.19 & 0.09 & 0.35 & 0.15\\ \hline
        PegDIG & 429 & 0.41 & 0.22 & 0.84 & 0.43\\ \hline
        Sculptor Dwarf Galaxy & 65 & 1.78 & 1.02 & 3.65 & 1.98\\ \hline
        Sextans dSph & 65 & 0.90 & 0.30 & 1.86 & 0.54\\ \hline
        Sgr dIG & 95 & 1.12 & 0.34 & 2.26 & 0.60\\ \hline
        UMi Galaxy & 44 & 1.89 & 0.81 & 3.79 & 1.50\\ \hline
        WLM Galaxy & 244 & 0.42 & 0.21 & 0.83 & 0.39\\ \hline
        Z 64-73 & 65 & 1.47 & 0.80 & 3.09 & 1.59\\ \hline
        Z 126-111 & 95 & 1.35 & 0.71 & 2.70 & 1.35\\ \hline
    \end{tabular}%
    }
    \label{tab:main-results}}
\end{table}
Here for each galaxy we specify the radius $\bar r$, at which we use the PSD to place the limit. For both of the methods we also give the value of sterile neutrino DM mass $\bar m$ which yields the best fit to the analysed stellar observational data with application of the GravSphere code. This ``best-fit mass'' is radius-dependent and is plotted for three example galaxies in Fig.\,\ref{fig:m-r-examples} with black solid line. The lower limits on DM mass are placed at 95\% CL. One observes that generally the EMF gives stronger limits than the maximum PSD. One also finds that while the best-fit mass varies from galaxy to galaxy within one order of magnitude the spread of 95\% CL limits is rather moderate. 

Only one galaxy, Hercules, provides the limit above 2\,keV, while five galaxies are present with limits above 1.5\,keV. Though we do not have independent reasons to explain why Hercules is an outlier, and can not say that there are specific problems with observations in this galaxy, we still prefer to be conservative, and will not use it for obtaining the lower bound on the DM mass. Thus, we choose the Sculptor galaxy to place our final constraints and conclude that PSD considerations give the lower limit
\[
m_{NRP}>1.98\,\text{keV} \;\;\;(95\%\,\text{CL})
\]
on sterile neutrino produced via oscillation in lepton-symmetric primordial plasma (Dodelson--Widrow mechanism \cite{Dodelson:1993je}). Note, that a quarter of our galaxy sample exhibits limits above 1.5\,keV. 

To compare the simple method of Sec.\,\ref{sec:analytical} and more consistent approach above we extract from the profiles, obtained with our GravSphere simulations, the velocity dispersions, half-light radii and their 2-sigma errors. Then we put these numbers into the analytic formulas of Sec.3 for $Q_{max}$ and into eq.\,\eqref{max-PSD-bound} for the mass limit. We observed that the central values of masses, obtained in this way, almost always exceed what we present in Table\,\ref{tab:main-results} and the 2 sigma limits are above the results in Table\,\ref{tab:main-results} for all the galaxies. We conclude that the consistent approach of this Section  generically gives more conservative estimates.

\section{Bounds in alternative models of sterile neutrino dark matter production}
\label{sec:alternatives}

The mechanism of sterile neutrino DM production we consider is the most simple, as it involves, apart from introduction to the SM of a new fermionic degree of freedom, only one parameter, mixing between sterile and active neutrinos. There are various extensions of this oscillation mechanism and others suggested in literature as responsible for the DM sterile neutrino production. The analysis of DM PSD in the dSphs performed in this paper can be applied to these models as well to constrain the DM mass and other parameters determining the PSD of the produced sterile neutrino component.  

To illustrate this possibility we consider here two models from the set considered in Ref.\,\cite{Gelmini:2019wfp}. Both models adopt the same mechanism of sterile neutrino production, that is  oscillations in the primordial lepton-symmetric plasma, but assume a non-standard cosmology. In the model I the Universe expansion is supposed to be dominated by the kinetic term of some scalar field. At this kination stage the dominant energy density drops with the scale factor $a$ as $\rho\propto a^{-6}$, much faster than the radiation or matter densities, and finally the Universe enters the radiation domination before the epoch of Big Bang Nucleosynthesis. The predicted spectrum of the sterile neutrino produced mostly at the kination stage reads 
\[
f_I(p/T_\nu)\propto \sqrt[3]{\frac{p}{T_\nu}}\l \frac{m}{\keV} \r^{2/3} \frac{1}{\text{e}^{p/T_\nu}+1}\,,
\]
and the normalisation coefficient is chosen to entirely explain the DM component. 

In the model II the Universe reheats very late, so the plasma emerges right before the Big Bang Nucleosynthesis, the largest temperature is supposed to be 5\,MeV. Once the plasma emerges, the sterile neutrino production in oscillations starts. However, it is strongly suppressed, since in the standard cosmology the sterile neutrinos dominantly produced at the temperature of about 150\,MeV. The resulting spectrum is approximated as 
\[
f_{II}(p/T_\nu)\propto \frac{p}{T_\nu} \frac{1}{\text{e}^{p/T_\nu}+1}\,,
\]
and the normalisation coefficient is tuned to form the whole DM component of the Universe. 

Our analysis of the PSD of 20 dSphs applied to the two models above yields the lower limits on the sterile neutrino DM presented in Tab.\,\ref{tab:alt-results}.  
\begin{table}[!htb]
\tbl{Limits on sterile neutrino DM mass in model with non-standard cosmology: model I with kination domination at sterile neutrino production and model II with very low energy reheating temperature, see main text for details. Then for both methods we present the central value of the sterile neutrino mass $\bar m$, with our formulas it provides the best fit to the observations. The columns $m>..$ contain the inferred lower limits at 95\% CL. All masses are in keV.}
{\centering
    \resizebox{\columnwidth}{!}{%
    \begin{tabular}{|l|c|c|c|c|c|c|c|c|}
    \hline
  \multicolumn{1}{|c|}{ Method $\rightarrow$} & 
  \multicolumn{4}{|c|}{Maximum PSD} & \multicolumn{4}{|c|}{EMF} \\
\hline 
 \multicolumn{1}{|c|}{ Model $\rightarrow$} & 
  \multicolumn{2}{|c|}{Model I} & 
 \multicolumn{2}{|c|}{Model II}  
 & \multicolumn{2}{|c|}{Model I} 
 & \multicolumn{2}{|c|}{Model II}  
 \\ \hline
Object $\downarrow$ & $\bar m$ & $m>..$ & $\bar m $ & $m>..$ & $\bar m$ & $m>..$ & $\bar m$ & $m>..$ \\\hline
     
        Andromeda V &  2.18 & 1.53 & 4.53 & 3.00 & 3.03 & 2.08 & 5.72 & 3.70 \\ \hline
        Aquarius Dwarf & 1.95 & 1.15 & 3.98 & 2.15 & 2.66 & 1.51 & 4.93 & 2.58 \\ \hline
        Bootes Dwarf Spheroidal Galaxy & 2.22 & 1.68 & 4.64 & 3.34 & 2.97 & 2.19 & 5.63 & 3.96 \\ \hline
        Carina dSph & 2.15 & 1.18 & 4.45 & 2.22 & 2.86 & 1.48 & 5.38 & 2.53 \\ \hline
        Cetus Dwarf Galaxy & 2.30 & 1.68 & 4.84 & 3.34 & 3.21 & 2.29 & 6.13 & 4.14 \\ \hline
        Coma Dwarf Galaxy & 1.39 & 1.31 & 2.68 & 2.50 & 1.74 & 1.63 & 3.05 & 2.84 \\ \hline
        CVn I dSph & 1.88 & 1.07 & 3.81 & 1.97 & 3.58 & 1.40 & 4.75 & 2.35 \\ \hline
        Dra dSph & 1.88 & 1.17 & 3.82 & 2.20 & 2.46 & 1.46 & 4.53 & 2.49 \\ \hline
        Fornax Dwarf Spheroidal & 1.67 & 1.17 & 3.32 & 2.20 & 2.27 & 1.57 & 4.09 & 2.67 \\ \hline
        Hercules Dwarf Galaxy & 3.99 & 2.17 & 9.20 & 4.52 & 5.64 & 2.94 & 11.76 & 5.55 \\ \hline
        Leo A & 2.23 & 1.61 & 4.66 & 3.19 & 3.00 & 2.11 & 5.69 & 3.79 \\ \hline
        NGC 6822 & 0.66 & 0.40 & 1.12 & 0.63 & 0.83 & 0.49 & 1.29 & 0.70 \\ \hline
        PegDIG & 1.07 & 0.72 & 1.97 & 1.24 & 1.46 & 0.96 & 2.45 & 1.51 \\ \hline
        Sculptor Dwarf Galaxy & 2.75 & 1.93 & 5.95 & 3.93 & 3.76 & 2.54 & 7.36 & 4.71 \\ \hline
        Sextans dSph & 1.78 & 0.87 & 3.58 & 1.55 & 2.44 & 1.10 & 4.45 & 1.78 \\ \hline
        Sgr dIG & 2.05 & 0.95 & 4.23 & 1.72 & 2.76 & 1.18 & 5.16 & 1.95 \\ \hline
        UMi Galaxy & 2.86  & 1.66 & 6.23 & 3.29 & 3.85 & 2.13 & 7.60 & 3.86 \\ \hline
        WLM Galaxy & 1.08 & 0.71 & 2.01 & 1.22 & 1.45 & 0.90 & 2.44 & 1.41 \\ \hline
        Z 64-73 & 2.44 & 1.65 & 5.17 & 3.27 & 3.37 & 2.20 & 6.47 & 3.98 \\ \hline
        Z 126-111 & 2.30 & 1.53 & 4.84 & 3.01 & 3.10 & 1.99 & 5.89 & 3.54 \\ \hline
    \end{tabular}%
    }
    \label{tab:alt-results}}
\end{table}
where we use the same notations as in 
Tab.\,\ref{tab:main-results}.  
This investigation confirms our observation that 
generally (both for galaxies and production mechanisms involved) the EMF gives stronger limits than maximum PSD does. Likewise, while the best-fit mass varies from galaxy to galaxy within one order of magnitude the scattering of 95\% CL limits is rather moderate. 

Analysis of the DM production in the model I with kination domination at production reveals only one galaxy, Hercules,  where we get the limits above 2.9\,keV, but eight  galaxies with limits above 2\,keV. So, being somewhat conservative we choose the Sculptor galaxy to obtain the lower limit from PSD considerations
\[
m_{NRP}>2.54\,\text{keV} \;\;\;(95\%\,\text{CL})
\]
on sterile neutrino produced via oscillation in lepton-symmetric primordial plasma (Dodelson--Widrow mechanism \cite{Dodelson:1993je}) in a model with kination domination down to the plasma temperature of $T_{reh}=5$\,MeV.  

Analysis of the DM production in the model II with low reheating temperature reveals only one galaxy where we get the limits above 5\,keV, but seven galaxies with limits above 3.5\,keV. Thus, again we choose the Sculptor galaxy to conclude that PSD considerations place the lower limit 
\[
m_{NRP}>4.71\,\text{keV} \;\;\;(95\%\,\text{CL})
\]
on sterile neutrino produced via oscillation in lepton-symmetric primordial plasma (Dodelson--Widrow mechanism \cite{Dodelson:1993je}) in a model with low reheating temperature $T_{reh}=5$\,MeV.

\section{Discussion}
\label{sec:Concl}

In this paper we analyse DM PSD in a set of dSphs and place a lower bound on sterile neutrino DM mass in three models of DM production in the early Universe. We use two quantities, maximum PSD and EMF, which only decrease during cosmic structure formation. Both can be estimated from observed characteristics of galaxy stars--their radial distribution and line-of-sight velocities--which implies averaging and consequent decrease of the inferred PSD. It allows us to place lower limits on the PSD parameters defined at the DM production in the early Universe.   

\begin{figure}[!htb]
    \centering
    \includegraphics[width=\linewidth]{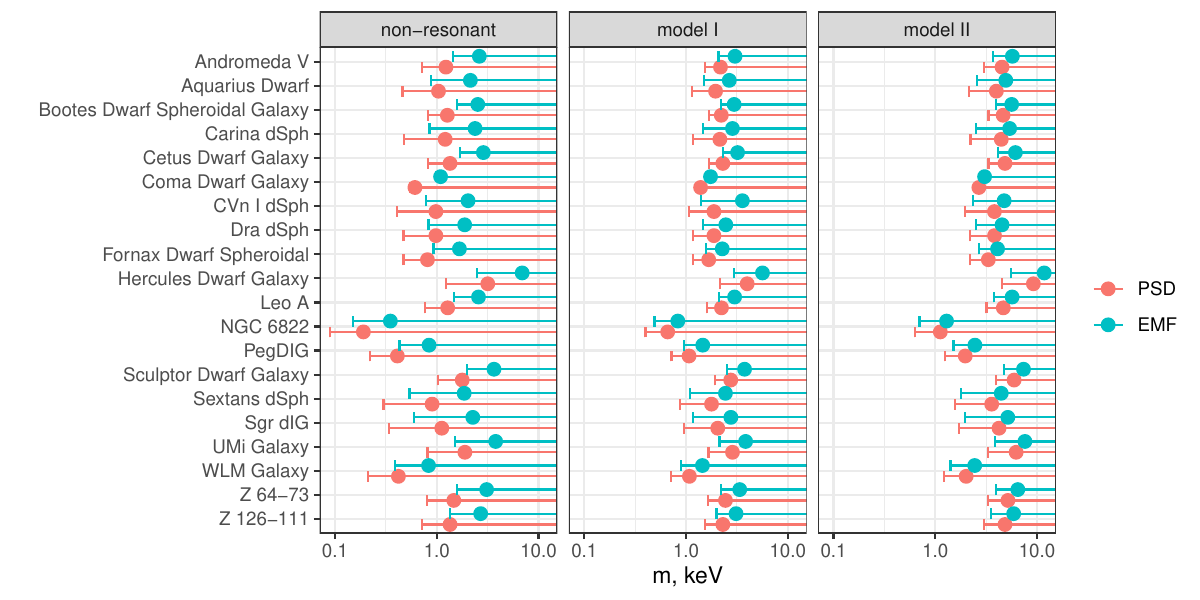}
    \caption{Summary of the bounds on the sterile neutrino DM obtained using maximum of PSD or EMF in the non-resonant production model in standard cosmology and in models I and II.}
    \label{fig:enter-label}
\end{figure}

We constrained sterile neutrino DM mass in models with production via active-sterile oscillations in the lepton-symmetric primordial plasma in cosmological models with standard cosmology, models with kination domination at production and models with very late reheating. The results based on the analysis of a set of the most promising (coldest and compact) dSphs are summarised in Fig.\,\ref{fig:enter-label}. 
One concludes that EMF always leads to stronger bounds than the maximum PSD. The method can be applied to other models of sterile neutrino DM production and in a wider range of WDM models, that is where the DM particle velocities are of the order of $10^{-3}$ at matter-radiation equality. 

To obtain the limits of Fig.\,\ref{fig:enter-label} we performed the GravSphere fit for a number of dSphs. The code utilizes MCMC technique to construct a set of models---galaxy DM profiles---which are statistically consistent with observed quantities: radial positions and line-of-sight velocities of the stars. The set of profiles allows us to estimate the total galaxy mass (mostly DM) enclosed inside a sphere of given radius and subsequently the stellar velocity dispersion. Assuming a model for DM velocity dispersion anisotropy one can obtain the DM PSD (which unfortunately can not be traced directly from the stellar observation data). Following Ref.\,\cite{Alvey:2020xsk} for the DM anisotropy we take the same parametrization, as for the stellar velocity anisotropy, but assume the parameters to be uniformly distributed within the prior range inspired by DM simulations (see Sec.\,\ref{sec:numerical}). There are two comments in order. 

First, for some of dSphs present in Tab.\,\ref{tab:galaxies-dataset}, a similar study was performed in Ref.\,\cite{Alvey:2020xsk} using the approach equivalent to our maximum PSD method, and our results do differ. Partly the difference is due to new stellar data and updated version of the GravShpere used in the present work. The version \cite{GravSphere_ref} used in Ref.\,\cite{Alvey:2020xsk}, adopts a piecewise power-law fit to the DM density profile with 5 radial bins, while we adopt the version of Refs.\,\cite{Read_2018,Collins_2021}, where a parametric analytic approximation for the DM density profile is used, based on a modified NFW profile with parametrically smoothed core and additional suppression at large radii. Another difference may be in the boundary conditions used for the solution to the Jeans equation \eqref{RJE DM} determining the galaxy DM velocity dispersion \eqref{sigma-DM-sol}. In our approach 
$\lim_{r\to\infty}\rho\sigma_r^2\to 0$, which is explicit in eq.\,\eqref{sigma-DM-sol}. Also, in our radial density fit the priors $[-5,\dots, -3.01]$ for the power $\alpha$ in the matter density asymptotic $\rho\propto r^\alpha$ at large radii $r\to\infty$ were used, which leadds to the finite galaxy mass. 
This was not the case in  Ref.\,\cite{Alvey:2020xsk}\footnote{The fits to DM velocity dispersion from \cite{Alvey:2020xsk} can be found at \url{https://github.com/james-alvey-42/FermionDSph/blob/master/Code/Final_Data/} in directories named after each galaxy in the files named \texttt{output\_sigr\_DM.pdf}.}. 

Second, our choice of the DM velocity anisotropy \eqref{fit-anis} is ad hoc to some extent, with the range of anisotropies selected as in Ref.~\citenum{Alvey:2020xsk}. The choice of Ref.~\citenum{Alvey:2020xsk} was inspired by comparison with EDGE \cite{Agertz_2019} and Aquarius \cite{Springel:2008cc} numerical simulations. We note, that choosing zero DM velocity anisotropy would lead to slightly stronger constraints, while increasing DM velocity anisotropy to the highest observed values of the stellar anisotropy would further weaken the constraints. Without possibility to measure DM velocities directly, a careful study of WDM structure formation could further define the proper choice of $\beta_{DM}$. Note, that recent analysis \cite{He:2024gvw} supports the use of non-zero $\beta_{DM}$ which is below stellar anisotropies at high radius. Note also, that choice of $\beta_{DM}=0$ would strengthen the obtained bounds by about 25\%.

In this respect, one needs more stellar data to improve the fit and tighten the bounds, or, at any rate, make them more robust. Recall that investigations of the PSD may reveal only lower bounds on the DM mass, not its value. In the absence of any systematics it may be tempting to take the strongest constraints from the 20 galaxies we used, which follow from the analysis of the Hercules dShp and read
\[
m>2.49\,\text{keV}\,,\;\;\;
m>2.94\,\text{keV}\,,\;\;\;
m>5.55\,\text{keV}\,\;\;\;
\]
for the sterile neutrinos within the three types of cosmology we considered.   However, a more prudent approach would be to consider Hercules to be an outlier due to unidentified observational systematic effects for this particular galaxy, and argue that weaker but more robust bounds are associated with the Sculptor galaxy 
\[
m>1.98\,\text{keV}\,,\;\;\;
m>2.54\,\text{keV}\,,\;\;\;
m>4.71\,\text{keV}\,\;\;\;
\]
respectively. Roughly one third of our set of galaxies gives very similar constraints to those from Sculptor, suggesting that this is a rather reliable observation. A statistical analysis of the combined results will be presented elsewhere. 

It is worth noting that models of WDM are typically constrained from observations of Cosmic Large Scale Structure, especially direct and indirect evidences for the existence of compact relatively light halos: dwarf galaxy counting in early and present Universe, Ly-$\alpha$ forest, gravitational lensing, possible perturbation of stellar streams, etc., for recent results see e.g.\ Refs.\,\cite{Villasenor:2022aiy,Dekker:2021scf,Montel:2022fhv,Banik:2019smi}. These limits depend on the spectrum of relic DM particles. Typically they are presented for the sterile neutrino non-resonant production, as we do, and also for the WDM thermal relic. The latter is supposed to be the situation where the DM particles decoupled in the early Universe from the relic plasma while being relativistic, so their spectrum is thermal. Also, it is assumed that a large entropy production happened in the late Universe, so the DM number density gets diluted and the final effective temperature of the DM component became much lower than the plasma temperature. In this case the present number density of DM particles is solely determined by its effective temperature, which can be defined from the the observed present DM abundance and DM particle mass. At the same time the DM velocity is proportional to $T_{effective}/m$. Recall that in the case of sterile neutrino DM we considered its velocity to be proportional to the ratio of the active neutrino temperature and sterile neutrino mass $T_\nu/m$, while overall distribution normalization is defined to reproduce the DM abundance. Hence, both models have two free parameters,  which are fixed by requirement of explaining the entire dark matter component and are constrained from the phase space density considerations. Therefore, the parameters are related to each other, and hence the sterile neutrino DM mass limit from maximal PSD can be converted to a limit on the mass $m_T$ of the thermal DM relics. One finds from eqs.\,\eqref{max-PSD-bound}, \eqref{add-1} and \eqref{Fermi max} the relation
\[
m_T=m\times \l \frac{5.58\,\text{eV}}{m}\r^{1/4}.
\]
Thus the above limits based on maximum PSD give 
\[
m_T>0.28\,\text{keV},
\]
while EMF limits translate to the following limit on the mass of the thermal DM
\[
m_T>0.49\,\text{keV}.
\]
 
Finally, there are bounds specific for the sterile neutrino DM originating from the search for the peak in the galaxy X-ray spectra possibly associated with the radiative decay of sterile neutrino galactic DM, for the most recent limits see Refs.\,\cite{Zakharov:2023mnp,Krivonos:2024yvm}. At a given sterile neutrino mass they place an upper bound on the sterile-active neutrino mixing. While the simple non-resonant production is excluded from these considerations, the phase space density provides with independent and complementary constraints on the sterile neutrino model parameters. At the same time, the alternative, resonant mechanism of sterile neutrino DM production\,\cite{Shi:1998km} is still viable\,\cite{Gorbunov:2025nqs}. The spectrum of the produced sterile neutrino DM is more complicated and dedicated the numerical simulations are needed to apply the phase space density constraints we suggest in this paper. This task is postponed to the future, while here we limited to simple cases of non-resonant production and thermal relic DM, allowing for simple comparison with other works.

\section*{Acknowledgements} 
We thank G.~Rubtsov and S.~Troitsky for valuable discussions on the applicability of statistical methods in astrophysics. 
The work is supported by the RSF grant 22-12-00271. E.K. acknowledges the Theoretical Physics and Mathematics Advancement Foundation 'BASIS'. 
Numerical calculations were performed on the Computational Cluster of the Theoretical division of INR RAS.

\bibliographystyle{ws-ijmpa}
\bibliography{refs_v2}

\begin{thebibliography}{10}
\expandafter\ifx\csname urlstyle\endcsname\relax
  \providecommand{\doi}[1]{doi:\discretionary{}{}{}#1}\else
  \providecommand{\doi}{doi:\discretionary{}{}{}\begingroup
  \urlstyle{rm}\Url}\fi

\bibitem{deVega:2009ku}
H.~J. de~Vega and N.~G. Sanchez, {\em Mon. Not. Roy. Astron. Soc.} {\bf 404},
  885  (2010), \href{http://arxiv.org/abs/0901.0922}{{\ttfamily arXiv:0901.0922
  [astro-ph.CO]}}, \doi{10.1111/j.1365-2966.2010.16319.x}.

\bibitem{Rajagopal:1990yx}
K.~Rajagopal, M.~S. Turner and F.~Wilczek, {\em Nucl. Phys. B} {\bf 358}, 447
  (1991), \doi{10.1016/0550-3213(91)90355-2}.

\bibitem{Gorbunov:2008ui}
D.~Gorbunov, A.~Khmelnitsky and V.~Rubakov, {\em JHEP} {\bf 12},   055  (2008),
  \href{http://arxiv.org/abs/0805.2836}{{\ttfamily arXiv:0805.2836 [hep-ph]}},
  \doi{10.1088/1126-6708/2008/12/055}.

\bibitem{King:2012wg}
S.~F. King and A.~Merle, {\em JCAP} {\bf 08},   016  (2012),
  \href{http://arxiv.org/abs/1205.0551}{{\ttfamily arXiv:1205.0551 [hep-ph]}},
  \doi{10.1088/1475-7516/2012/08/016}.

\bibitem{Kusenko:2009up}
A.~Kusenko, {\em Phys. Rept.} {\bf 481}, 1  (2009),
  \href{http://arxiv.org/abs/0906.2968}{{\ttfamily arXiv:0906.2968 [hep-ph]}},
  \doi{10.1016/j.physrep.2009.07.004}.

\bibitem{Merle:2013gea}
A.~Merle, {\em Int. J. Mod. Phys. D} {\bf 22},   1330020  (2013),
  \href{http://arxiv.org/abs/1302.2625}{{\ttfamily arXiv:1302.2625 [hep-ph]}},
  \doi{10.1142/S0218271813300206}.

\bibitem{Drewes:2016upu}
M.~Drewes {\em et~al.}, {\em JCAP} {\bf 01},   025  (2017),
  \href{http://arxiv.org/abs/1602.04816}{{\ttfamily arXiv:1602.04816
  [hep-ph]}}, \doi{10.1088/1475-7516/2017/01/025}.

\bibitem{Abazajian:2017tcc}
K.~N. Abazajian, {\em Phys. Rept.} {\bf 711-712}, 1  (2017),
  \href{http://arxiv.org/abs/1705.01837}{{\ttfamily arXiv:1705.01837
  [hep-ph]}}, \doi{10.1016/j.physrep.2017.10.003}.

\bibitem{Boyarsky:2018tvu}
A.~Boyarsky, M.~Drewes, T.~Lasserre, S.~Mertens and O.~Ruchayskiy, {\em Prog.
  Part. Nucl. Phys.} {\bf 104}, 1  (2019),
  \href{http://arxiv.org/abs/1807.07938}{{\ttfamily arXiv:1807.07938
  [hep-ph]}}, \doi{10.1016/j.ppnp.2018.07.004}.

\bibitem{Boyarsky:2008ju}
A.~Boyarsky, O.~Ruchayskiy and D.~Iakubovskyi, {\em JCAP} {\bf 03},   005
  (2009), \href{http://arxiv.org/abs/0808.3902}{{\ttfamily arXiv:0808.3902
  [hep-ph]}}, \doi{10.1088/1475-7516/2009/03/005}.

\bibitem{Gorbunov:2008ka}
D.~Gorbunov, A.~Khmelnitsky and V.~Rubakov, {\em JCAP} {\bf 10},   041  (2008),
  \href{http://arxiv.org/abs/0808.3910}{{\ttfamily arXiv:0808.3910 [hep-ph]}},
  \doi{10.1088/1475-7516/2008/10/041}.

\bibitem{Horiuchi:2013noa}
S.~Horiuchi, P.~J. Humphrey, J.~Onorbe, K.~N. Abazajian, M.~Kaplinghat and
  S.~Garrison-Kimmel, {\em Phys. Rev. D} {\bf 89},   025017  (2014),
  \href{http://arxiv.org/abs/1311.0282}{{\ttfamily arXiv:1311.0282
  [astro-ph.CO]}}, \doi{10.1103/PhysRevD.89.025017}.

\bibitem{Wang:2017hof}
M.-Y. Wang, J.~F. Cherry, S.~Horiuchi and L.~E. Strigari (12 2017),
  \href{http://arxiv.org/abs/1712.04597}{{\ttfamily arXiv:1712.04597
  [astro-ph.CO]}}.

\bibitem{Alvey:2020xsk}
J.~Alvey, N.~Sabti, V.~Tiki, D.~Blas, K.~Bondarenko, A.~Boyarsky, M.~Escudero,
  M.~Fairbairn, M.~Orkney and J.~I. Read, {\em Mon. Not. Roy. Astron. Soc.}
  {\bf 501}, 1188  (2021), \href{http://arxiv.org/abs/2010.03572}{{\ttfamily
  arXiv:2010.03572 [hep-ph]}}, \doi{10.1093/mnras/staa3640}.

\bibitem{Kaplinghat:2005sy}
M.~Kaplinghat, {\em Phys. Rev. D} {\bf 72},   063510  (2005),
  \href{http://arxiv.org/abs/astro-ph/0507300}{{\ttfamily
  arXiv:astro-ph/0507300}}, \doi{10.1103/PhysRevD.72.063510}.

\bibitem{Munoz:2018}
R.~R. {Mu{\~n}oz}, P.~{C{\^o}t{\'e}}, F.~A. {Santana}, M.~{Geha}, J.~D.
  {Simon}, G.~A. {Oyarz{\'u}n}, P.~B. {Stetson} and S.~G. {Djorgovski}, {\em
  Astrophys. J.} {\bf 860},  ~66  (2018),
  \href{http://arxiv.org/abs/1806.06891}{{\ttfamily arXiv:1806.06891
  [astro-ph.GA]}}, \doi{10.3847/1538-4357/aac16b}.

\bibitem{Dodelson:1993je}
S.~Dodelson and L.~M. Widrow, {\em Phys. Rev. Lett.} {\bf 72}, 17  (1994),
  \href{http://arxiv.org/abs/hep-ph/9303287}{{\ttfamily arXiv:hep-ph/9303287}},
  \doi{10.1103/PhysRevLett.72.17}.

\bibitem{Gelmini:2019wfp}
G.~B. Gelmini, P.~Lu and V.~Takhistov, {\em JCAP} {\bf 12},   047  (2019),
  \href{http://arxiv.org/abs/1909.13328}{{\ttfamily arXiv:1909.13328
  [hep-ph]}}, \doi{10.1088/1475-7516/2019/12/047}.

\bibitem{Lynden-Bell:1966zjv}
D.~Lynden-Bell, {\em Mon. Not. Roy. Astron. Soc.} {\bf 136}, 101  (1967).

\bibitem{STremaine}
S.~Tremaine, M.~Henon and D.~Lynden-Bell, {\em Mon. Not. Roy. Astron. Soc.}
  {\bf 285}, 219  (1986).

\bibitem{1986AJ.....92...72R}
D.~O. Richstone and S.~Tremaine, {\em Astronom. J.} {\bf 92}, 72  (1986),
  \doi{10.1086/114135}.

\bibitem{GravSphere_ref}
J.~I. Read and P.~Steger, {\em Mon. Not. Roy. Astron. Soc.} {\bf 471}, 4541
  (2017), \href{http://arxiv.org/abs/1701.04833}{{\ttfamily arXiv:1701.04833
  [astro-ph]}}, \doi{10.1093/mnras/stx1798}.

\bibitem{Read_2018}
J.~I. Read, M.~G. Walker and P.~Steger, {\em Mon. Not. Roy. Astron. Soc.} {\bf
  481},   860–877  (2018), \href{http://arxiv.org/abs/1805.06934}{{\ttfamily
  arXiv:1805.06934 [astro-ph.GA]}}, \doi{10.1093/mnras/sty2286}.

\bibitem{Collins_2021}
M.~L.~M. Collins {\em et~al.}, {\em Mon. Not. Roy. Astron. Soc.} {\bf 505},
  5686–5701  (2021), \href{http://arxiv.org/abs/2102.11890}{{\ttfamily
  arXiv:2102.11890 [astro-ph.GA]}}, \doi{10.1093/mnras/stab1624}.

\bibitem{MCMC_ref}
D.~Foreman-Mackey, D.~W. Hogg, D.~Lang and J.~Goodman, {\em Publ. Astron. Soc.
  Pac.} {\bf 125},   306–312  (2013),
  \href{http://arxiv.org/abs/1202.3665}{{\ttfamily arXiv:1202.3665
  [astro-ph.IM]}}, \doi{10.1086/670067}.

\bibitem{2012AJ....144....4M}
A.~W. McConnachie, {\em Astrophys. J.} {\bf 144},  ~4  (2012),
  \href{http://arxiv.org/abs/1204.1562}{{\ttfamily arXiv:1204.1562
  [astro-ph.CO]}}, \doi{10.1088/0004-6256/144/1/4}.

\bibitem{Genina_2020}
A.~Genina {\em et~al.}, {\em Mon. Not. Roy. Astron. Soc.} {\bf 498},
  144–163  (2020), \href{http://arxiv.org/abs/1911.09124}{{\ttfamily
  arXiv:1911.09124 [astro-ph.GA]}}, \doi{10.1093/mnras/staa2352}.

\bibitem{2018MNRAS.479.4136K}
I.~D. Karachentsev, E.~I. Kaisina and D.~I. Makarov, {\em Mon. Not. Roy.
  Astron. Soc.} {\bf 479},   4136–4152  (2018),
  \href{http://arxiv.org/abs/1806.09822}{{\ttfamily arXiv:1806.09822
  [astro-ph.GA]}}, \doi{10.1093/mnras/sty1774}.

\bibitem{2011ApJS..192....6L}
J.~C. Lee, A.~Gil~de Paz, R.~C. Kennicutt, M.~Bothwell, J.~Dalcanton, J.~G.~F.
  S.~J., B.~D. Johnson, S.~Sakai, E.~Skillman, C.~Tremonti and L.~van Zee, {\em
  ApJS} {\bf 192},  ~6  (2010),
  \href{http://arxiv.org/abs/1009.4705}{{\ttfamily arXiv:1009.4705
  [astro-ph.CO]}}, \doi{10.1088/0067-0049/192/1/6}.

\bibitem{2014MNRAS.445..881C}
D.~O. Cook, D.~A. Dale, B.~D. Johnson, L.~Van~Zee, J.~C. Lee, R.~C. Kennicutt,
  D.~Calzetti, S.~M. Staudaher and C.~W. Engelbracht, {\em Mon. Not. Roy.
  Astron. Soc.} {\bf 445},   881–889  (2014),
  \href{http://arxiv.org/abs/1408.1130}{{\ttfamily arXiv:1408.1130
  [astro-ph.GA]}}, \doi{10.1093/mnras/stu1580}.

\bibitem{2020ApJ...893...47D}
A.~Drlica-Wagner {\em et~al.}, {\em Astrophys. J.} {\bf 893},  ~47  (2020),
  \href{http://arxiv.org/abs/1912.03302}{{\ttfamily arXiv:1912.03302
  [astro-ph.GA]}}, \doi{10.3847/1538-4357/ab7eb9}.

\bibitem{2018ApJ...861...49H}
M.~P. Haynes {\em et~al.}, {\em Astrophys. J.} {\bf 861},  ~49  (2018),
  \href{http://arxiv.org/abs/1912.03302}{{\ttfamily arXiv:1912.03302
  [astro-ph.GA]}}, \doi{10.3847/1538-4357/aac956}.

\bibitem{Koreshk_github}
\url{https://github.com/koreshk/Estimation-of-phase-space-density-in-dwarf-galaxies}.

\bibitem{Agertz_2019}
O.~Agertz, A.~Pontzen, J.~I. Read, M.~P. Rey, M.~Orkney, J.~Rosdahl,
  R.~Teyssier, R.~Verbeke, M.~Kretschmer and S.~Nickerson, {\em Mon. Not. Roy.
  Astron. Soc.} {\bf 491}, 1656–  (2019),
  \href{http://arxiv.org/abs/1904.02723}{{\ttfamily arXiv:1904.02723
  [astro-ph.GA]}}, \doi{10.1093/mnras/stz3053}.

\bibitem{Springel:2008cc}
V.~Springel, J.~Wang, M.~Vogelsberger, A.~Ludlow, A.~Jenkins, A.~Helmi, J.~F.
  Navarro, C.~S. Frenk and S.~D.~M. White, {\em Mon. Not. Roy. Astron. Soc.}
  {\bf 391}, 1685  (2008), \href{http://arxiv.org/abs/0809.0898}{{\ttfamily
  arXiv:0809.0898 [astro-ph]}}, \doi{10.1111/j.1365-2966.2008.14066.x}.

\bibitem{He:2024gvw}
J.~He {\em et~al.}, {\em Astrophys. J.} {\bf 976},   187  (2024),
  \href{http://arxiv.org/abs/2407.14827}{{\ttfamily arXiv:2407.14827
  [astro-ph.GA]}}, \doi{10.3847/1538-4357/ad8882}.

\bibitem{Villasenor:2022aiy}
B.~Villasenor, B.~Robertson, P.~Madau and E.~Schneider, {\em Phys. Rev. D} {\bf
  108},   023502  (2023), \href{http://arxiv.org/abs/2209.14220}{{\ttfamily
  arXiv:2209.14220 [astro-ph.CO]}}, \doi{10.1103/PhysRevD.108.023502}.

\bibitem{Dekker:2021scf}
A.~Dekker, S.~Ando, C.~A. Correa and K.~C.~Y. Ng, {\em Phys. Rev. D} {\bf 106},
    123026  (2022), \href{http://arxiv.org/abs/2111.13137}{{\ttfamily
  arXiv:2111.13137 [astro-ph.CO]}}, \doi{10.1103/PhysRevD.106.123026}.

\bibitem{Montel:2022fhv}
N.~A. Montel, A.~Coogan, C.~Correa, K.~Karchev and C.~Weniger, {\em Mon. Not.
  Roy. Astron. Soc.} {\bf 518}, 2746  (2022),
  \href{http://arxiv.org/abs/2205.09126}{{\ttfamily arXiv:2205.09126
  [astro-ph.CO]}}, \doi{10.1093/mnras/stac3215}.

\bibitem{Banik:2019smi}
N.~Banik, J.~Bovy, G.~Bertone, D.~Erkal and T.~J.~L. de~Boer, {\em JCAP} {\bf
  10},   043  (2021), \href{http://arxiv.org/abs/1911.02663}{{\ttfamily
  arXiv:1911.02663 [astro-ph.GA]}}, \doi{10.1088/1475-7516/2021/10/043}.

\bibitem{Zakharov:2023mnp}
E.~I. Zakharov {\em et~al.}, {\em Phys. Rev. D} {\bf 109},   L021301  (2024),
  \href{http://arxiv.org/abs/2303.12673}{{\ttfamily arXiv:2303.12673
  [astro-ph.HE]}}, \doi{10.1103/PhysRevD.109.L021301}.

\bibitem{Krivonos:2024yvm}
R.~A. Krivonos, V.~V. Barinov, A.~A. Mukhin and D.~S. Gorbunov, {\em Phys. Rev.
  Lett.} {\bf 133},   261002  (2024),
  \href{http://arxiv.org/abs/2405.17861}{{\ttfamily arXiv:2405.17861
  [hep-ph]}}, \doi{10.1103/PhysRevLett.133.261002}.

\bibitem{Shi:1998km}
X.-D. Shi and G.~M. Fuller, {\em Phys. Rev. Lett.} {\bf 82}, 2832  (1999),
  \href{http://arxiv.org/abs/astro-ph/9810076}{{\ttfamily
  arXiv:astro-ph/9810076}}, \doi{10.1103/PhysRevLett.82.2832}.

\bibitem{Gorbunov:2025nqs}
D.~Gorbunov, D.~Kalashnikov and G.~Krugan (2 2025),
  \href{http://arxiv.org/abs/2502.17374}{{\ttfamily arXiv:2502.17374
  [hep-ph]}}.

\end{thebibliography}

\end{document}